\documentclass[a4paper,11pt,table]{article}
\pdfoutput=1 

\usepackage{jheppub} 

\usepackage[T1]{fontenc} 
\usepackage[utf8]{inputenc} 
\usepackage{microtype} 
\usepackage{lmodern} 
\usepackage{booktabs} 
\usepackage{mdframed} 
\usepackage{graphicx}
\usepackage{todonotes} 
\usepackage{slashed}
\usepackage[utf8]{inputenc}
\usepackage{cancel}

\usepackage{caption}
\usepackage{array}   
\newcolumntype{L}{>{$}l<{$}}
\newcolumntype{C}{>{$}c<{$}}

\usepackage{hyperref}
 \usepackage{color}
 \definecolor{dark-red}{rgb}{0.4,0.15,0.15}
 \definecolor{dark-blue}{rgb}{0.15,0.15,0.4}
 \definecolor{medium-blue}{rgb}{0,0,0.5}
 \hypersetup{colorlinks, linkcolor={dark-red}, citecolor={dark-blue}, urlcolor={medium-blue}}

\usepackage{amsmath,amsfonts,amssymb}
\usepackage{eucal} 
\usepackage{mleftright} \mleftright
\usepackage{tensor} 
\usepackage{braket} 
\usepackage{mathrsfs} 
\usepackage{array}

\providecommand*{\dd}{\mathop{}\!d}
\renewcommand*{\dd}{\mathop{}\!d}

\providecommand*{\vd}{\mathop{}\!\delta}
\renewcommand*{\vd}{\mathop{}\!\delta}

\providecommand*{\R}{{\mathbb{R}}}
\renewcommand*{\R}{{\mathbb{R}}}
\providecommand*{\C}{{\mathbb{C}}}
\renewcommand*{\C}{{\mathbb{C}}}
\providecommand*{\Sp}{{\mathbb{S}}}
\renewcommand*{\Sp}{{\mathbb{S}}}
\providecommand*{\Hy}{{\mathbb{H}}}
\renewcommand*{\Hy}{{\mathbb{H}}}

\providecommand*{\eqex}{\overset{!}{=}}
\renewcommand*{\eqex}{\overset{!}{=}}

\providecommand*{\dis}{\oplus}
\renewcommand*{\dis}{\oplus}

\usepackage{tikz}
\usepackage{tikz-cd}
\usepackage{tikz-3dplot}

\newcommand{\shi}{5}

\newcommand\parmp{\mathbin{\vcenter{\hbox{%
  \oalign{$\scriptstyle({-})$\cr
          \noalign{\kern-0.8ex}
          \hfil$\scriptscriptstyle+$\hfil\cr}%
}}}}

\providecommand{\Gt}{{\mathtt{G}}}
\renewcommand{\Gt}{{\mathtt{G}}}
\providecommand{\Bt}{{\mathtt{B}}}
\renewcommand{\Bt}{{\mathtt{B}}}
\providecommand{\Jt}{{\mathtt{J}}}
\renewcommand{\Jt}{{\mathtt{J}}}
\providecommand{\Ht}{{\mathtt{H}}}
\renewcommand{\Ht}{{\mathtt{H}}}
\providecommand{\St}{{\mathtt{S}}}
\renewcommand{\St}{{\mathtt{S}}}

\providecommand{\Pt}{{\mathtt{P}}}
\renewcommand{\Pt}{{\mathtt{P}}}
\providecommand{\Lt}{{\mathtt{L}}}
\renewcommand{\Lt}{{\mathtt{L}}}

\providecommand{\Zt}{{\mathtt{Z}}}
\renewcommand{\Zt}{{\mathtt{Z}}}
\providecommand{\Mt}{{\mathtt{M}}}
\renewcommand{\Mt}{{\mathtt{M}}}

\providecommand{\Mt}{{\mathtt{M}}}
\renewcommand{\Mt}{{\mathtt{M}}}
\providecommand{\Xt}{{\mathtt{X}}}
\renewcommand{\Xt}{{\mathtt{X}}}
\providecommand{\Yt}{{\mathtt{Y}}}
\renewcommand{\Yt}{{\mathtt{Y}}}

\providecommand{\cK}{{\mathcal{K}}}
\renewcommand{\cK}{{\mathcal{K}}}

\providecommand{\cM}{{\mathcal{M}}}
\renewcommand{\cM}{{\mathcal{M}}}

\providecommand{\fm}{{\mathfrak{m}}}
\renewcommand{\fm}{{\mathfrak{m}}}
\providecommand{\fk}{{\mathfrak{k}}}
\renewcommand{\fk}{{\mathfrak{k}}}
\providecommand{\fh}{{\mathfrak{h}}}
\renewcommand{\fh}{{\mathfrak{h}}}
\providecommand{\fg}{{\mathfrak{g}}}
\renewcommand{\fg}{{\mathfrak{g}}}

\providecommand{\homoh}{{\mathfrak{b}}}
\renewcommand{\homoh}{{\mathfrak{b}}}
\providecommand{\homoH}{{\mathcal{B}}}
\renewcommand{\homoH}{{\mathcal{B}}}

\newcommand{\hJ}{\hat{{\mathtt{J}}}}
\newcommand{\hP}{\hat{{\mathtt{P}}}}

\providecommand{\el}{\Lambda}
\renewcommand{\el}{\Lambda}
\providecommand{\ec}{c_{i}}
\renewcommand{\ec}{c_{i}}
\providecommand{\et}{\tau}
\renewcommand{\et}{\tau}

\providecommand*{\D}{{\natural}}
\renewcommand*{\D}{{\natural}}

\providecommand{\mugr}{\mu}
\renewcommand{\mugr}{\mu}
\providecommand{\muex}{\chi}
\renewcommand{\muex}{\chi}

\usepackage{pifont}
\newcommand{\cmark}{{\LARGE \ding{51}}}
\newcommand{\xmark}{{\LARGE \ding{55}}}
\newcommand{\cm}{{\ding{51}}}
\newcommand{\xm}{{\ding{55}}}

\newcommand{\yes}{{\text{\ding{51}}}}
\newcommand{\no}{{\text{\ding{55}}}}

\usepackage{amsthm} 
\theoremstyle{plain}
\newtheorem{theorem}{Theorem}

\theoremstyle{definition}

\theoremstyle{remark}

\usepackage{float}

\title{\boldmath Limits of three-dimensional gravity and metric
  kinematical Lie algebras in any dimension}

\author[a,1]{Javier Matulich\note{\href{https://orcid.org/0000-0002-3558-9025}{ORCID: 0000-0002-3558-9025}}}
\author[a,2]{Stefan Prohazka\note{\href{https://orcid.org/0000-0002-3925-3983}{ORCID: 0000-0002-3925-3983}}}
\author[b,3]{and Jakob Salzer\note{\href{https://orcid.org/0000-0002-9560-344X}{ORCID: 0000-0002-9560-344X}}}

\affiliation[a]{Universit\'e Libre de Bruxelles and International Solvay Institutes,\\
  Physique Math\'ematique des Interactions Fondamentales,\\
  Campus Plaine---CP~231,
  Bruxelles B-1050, Belgium}
\affiliation[b]{Departament  de  Física  Quàntica  i  Astrofísica,  Institut  de  Ciències  del  Cosmos,\\
Universitat  de  Barcelona,  Martí  i  Franquès  1,
E-08028  Barcelona,  Spain}

\emailAdd{javier.matulich@ulb.ac.be}
\emailAdd{stefan.prohazka@ulb.ac.be}
\emailAdd{jakob.salzer@icc.ub.edu}

\abstract{We extend a recent classification of three-dimensional
  spatially isotropic homogeneous spacetimes to Chern--Simons theories
  as three-dimensional gravity theories on these spacetimes. By this
  we find gravitational theories for all carrollian, galilean, and
  aristotelian counterparts of the lorentzian theories. In order to
  define a nondegenerate bilinear form for each of the theories, we
  introduce (not necessarily central) extensions of the original
  kinematical algebras. Using the structure of so-called double
  extensions, this can be done systematically. For homogeneous spaces
  that arise as a limit of (anti-)de Sitter spacetime, we show that it
  is possible to take the limit on the level of the action, after an
  appropriate extension. We extend our systematic construction of
  nondegenerate bilinear forms also to all higher-dimensional
  kinematical algebras.}

\dedicated{Christa Zauner gewidmet.}

\begin{document} 
\maketitle
\flushbottom

\section{Introduction}
\label{sec:introduction}

The crucial rôle played by symmetries in three-dimensional gravity
becomes obvious from the intriguing possibility to write gravity in a
gauge theory formulation. The Chern--Simons (CS) action based on
either of the gauge algebras
$\mathfrak{iso}(2,1), \mathfrak{so}(3,1), \mathfrak{so}(2,2)$, is
classically equivalent to the Einstein--Hilbert action in its
first-order formulation with zero, positive or negative cosmological
constant, respectively \cite{Achucarro:1987vz,Witten:1988hc}. Since
three-dimensional gravity does not allow for propagating degrees of
freedom, the solutions of the respective theories are consequently
locally three-dimensional Minkowski, de Sitter (dS), or anti-de Sitter
(AdS), with the gauge algebras corresponding to the symmetries of the
respective spacetime. Gravity on $\textrm{AdS}_3$ in particular has
received a great amount of attention as one finds that, imposing the
right boundary conditions \cite{Brown:1986nw}, the asymptotic
symmetries of $\textrm{AdS}_3$ yield the infinite-dimensional
symmetries of a two-dimensional conformal field theory (CFT). This
set-up presents one of the best studied instances of the AdS/CFT
duality \cite{Maldacena:1997re,Gubser:1998bc,Witten:1998qj}. The
intimate relation of CS theories to Wess--Zumino--Witten (WZW) models
\cite{Witten:1988hf,Elitzur:1989nr} is a reason the gauge formulation
of $\textrm{AdS}_3$ gravity appears to be particularly suited for a
better understanding of the correspondence
\cite{Coussaert:1995zp}.

The above mentioned symmetry algebras, $\mathfrak{iso}(2,1)$,
$\mathfrak{so}(3,1)$, $\mathfrak{so}(2,2)$, are examples of
\emph{kinematical Lie algebras}, that are associated to
three-dimensional \emph{homogeneous spacetimes} on which they act
transitively; for the above mentioned these spacetimes are precisely
three-dimensional Minkowski, $\textrm{dS}_3$, and $\textrm{AdS}_3$,
respectively.\footnote{For precise definitions and further details see
  Appendix~\ref{sec:kinbackground}.} From the action on the
homogeneous spacetime the generators of a kinematical algebra inherit
an interpretation of rotations, boosts, spatial and temporal
translations. In fact, any homogeneous spacetime is completely
specified by a kinematical algebra $\mathfrak{k}$ and a choice of
subalgebra $\homoh$ that singles out the transformations associated to
boosts and rotations. For instance, for all three symmetry algebras
mentioned above, the subalgebra $\homoh$ is the three-dimensional
Lorentz algebra $\mathfrak{so}(2,1)$. The pair $(\mathfrak{k},\homoh)$
determining the homogeneous space is called \emph{Klein pair}.
Assuming $\cK$ and $\homoH$ are the simply connected Lie groups with
Lie algebra $\fk$ and $\homoh$, respectively, the homogeneous
spacetime is then given by the quotient $\cK/\homoH$.

In this work we want to define CS theories for \emph{all possible
  three-dimensional spatially isotropic homogeneous spacetimes}. These
homogeneous spacetimes can be regarded as generalization of maximally
symmetric spacetimes dropping the assumption of the existence of a
nondegenerate tangent space metric. The resulting CS theories are
consequently gravitational theories that may not allow for a
description in terms of a metric, i.e., they are non- or
ultrarelativistic theories of gravity that have instead of the metric
a \emph{galilean} or \emph{carrollian} structure on the tangent space.

The interest in gravitational theories based on galilean symmetries
has been rekindled recently due to their relevance for holographic
correspondences
\cite{Son:2008ye,Balasubramanian:2008dm,Kachru:2008yh,Bagchi:2009my,Bagchi:2009pe,Christensen:2013lma,Christensen:2013rfa,Taylor:2015glc}
and effective field theories
\cite{Son:2013rqa,Hoyos:2011ez,Gromov:2015fda,Geracie:2015dea} for
condensed matter systems. Ultrarelativistic gravity theories on the
other hand could help to elucidate the structure of field theories
with carrollian symmetries, which appear at event horizons
\cite{Blau:2015nee,Penna:2018gfx} and at null infinity (and arguably
also at spatial infinity
\cite{Troessaert:2017jcm,Henneaux:2018cst,Gibbons:2019zfs}).\footnote{More
  precisely: The structure at null infinity is only a \emph{conformal
    carrollian structure}. This structure is left invariant by the
  infinite-dimensional Bondi--van der Burg--Metzner--Sachs (BMS) group
  \cite{Bondi:1962px,Sachs:1962zza,Duval:2014uva}.} Given the
simplicity of gravity in three dimensions, it suggests itself as an
interesting testing ground for the study of gravitational theories
with non-lorentzian symmetries; see
\cite{Hartong:2015xda,Bergshoeff:2016soe,Bergshoeff:2017btm} for CS
theories of the Carroll group and
\cite{Papageorgiou:2009zc,Papageorgiou:2010ud,Bergshoeff:2016lwr,Hartong:2016yrf,Joung:2018frr}
for works on the (AdS-)Galilei group and supersymmetric extensions
thereof.

A CS theory for a homogeneous spacetime is defined by two ingredients
\begin{itemize}
\item a Klein pair $(\fk,\homoh)$ with $\fk$ a kinematical Lie algebra
  and $\homoh$ a subalgebra determining boosts and rotations;
\item a symmetric, invariant, nondegenerate bilinear form, i.e., an \emph{invariant metric} on $\fk$.
\end{itemize}
In order to write down a CS theory for all three-dimensional
homogeneous spacetimes one has to, firstly, classify all Klein pairs
giving rise to a three-dimensional spacetime and, secondly, specify
an invariant metric for all of them.

A first classification of kinematical Lie algebras and their
associated homogeneous spacetimes was provided in a seminal paper by
Bacry and Lévy-Leblond \cite{Bacry:1968zf} (see also
\cite{Bacry:1986pm}) where they showed how these algebras can be
obtained by successive İnönü--Wigner (IW) contractions starting from
the symmetry algebra $\mathfrak{so}(D+1,1)$ [$\mathfrak{so}(D,2)$] of
$(D+1)$-dimensional [anti-]de Sitter space.\footnote{The authors of
  \cite{Bacry:1968zf} restricted to $3+1$ dimensions. It was only
  recently shown that there is no change for higher
  dimensions~\cite{Figueroa-OFarrill:2018ilb} (a property that is not
  true if solely the kinematical Lie algebras are
  considered~\cite{Figueroa-OFarrill:2017sfs}).} In this way one
recovers the well-known Poincaré algebra $\mathfrak{iso}(D,1)$, the
nonrelativistic galilean algebras, and the ultrarelativistic
carrollian algebras. In summary, they find a cube of kinematical
algebras connected by limiting procedures. One of the main results of
the present work is a lift of the cube of kinematical algebras of
Bacry and Lévy-Leblond to a cube of three-dimensional gravity theories
(cf.\ Figure \ref{fig:cube}) connected by various limits.

More recently, the classification of homogeneous spacetimes was
generalized to include spacetimes that do not arise as limits from
(anti-)de Sitter space as well as aristotelian spacetimes, which lack
boost invariance \cite{Figueroa-OFarrill:2018ilb}. We include these
\emph{non-contracting spacetimes} in our discussion of all possible
three-dimensional CS theories.

Let us now turn to the second ingredient in the definition of a CS
theory. The existence of an invariant metric is guaranteed for
[semi]simple Lie algebras such as the [A]dS algebra
$\mathfrak{so}(3,1)$ [$\mathfrak{so}(2,2)$]. However, apart from the
(A)dS algebra none of the other kinematical algebras appearing in the
cube of Bacry--Lévy-Leblond are semisimple. In line with this, one
finds that the galilean algebras do not allow for an invariant metric.
Yet the three-dimensional Poincaré algebra $\mathfrak{iso}(2,1)$ while
not being semisimple allows for an invariant metric, in contrast to
all higher dimensions; a fact that was crucial in the construction
of~\cite{Witten:1988hc}.

This raises the question which Lie algebras (or, for our purpose,
kinematical Lie algebras) allow for an invariant metric. The answer is
provided by a theorem of Medina and Revoy
\cite{Medina1985,FigueroaO'Farrill:1995cy} stating that all such Lie
algebras are either abelian, simple, or so-called \emph{double
  extensions} (or direct products of these ingredients). We will show
that the Poincaré algebra allows for an invariant metric precisely
because it can be understood as a double extension.

Given the fact that the galilean algebras do not allow for an
invariant metric it appears that our program to define a CS theory for
all three-dimensional homogeneous spacetimes is already doomed to
fail. However, the theorem of Medina and Revoy also provides us with
the tools to circumvent this conclusion. We find that the double
extension structure can be used as a bootstrap to construct out of any
kinematical algebra without invariant metric a new extended algebra
that has one by construction. The new algebra and the original
kinematical algebra share the same homogeneous spacetime. However,
this extension might not be unique, and one of the central goals of
this work is to find the one with \emph{minimal} additional
generators.

With this tool at our disposal, together with the classification of
all three-dimensional kinematical spacetimes
\cite{Figueroa-OFarrill:2018ilb}, we arrive at our main result which
is a classification of \emph{all Chern--Simons theories based on the
  gauging of three-dimensional kinematical Klein pairs}.

As mentioned above, we will also show that \emph{all} CS actions based
on the kinematical algebras of the cube of Bacry--Lévy-Leblond can be
obtained as limits of the trivially twice centrally extended (A)dS
algebra. These central extensions become nontrivial after the IW
contraction and lead to kinematical algebras having the form of double
extensions with invariant metric.\footnote{%
  In order to avoid possible confusion we want to stress that, despite
  the similar name, a double extension is a priori completely
  unrelated to a Lie algebra with two central extension.} A similar
approach for the case of the galilean algebras was pioneered in
\cite{Papageorgiou:2009zc,Papageorgiou:2010ud} and has recently been
revived~\cite{Bergshoeff:2016lwr,Hartong:2016yrf}. Remarkably these
limits are not only a limit of the equations of motion, but also
generalize to the action principle, which might be crucial for going
beyond the classical level.

Having discussed three-dimensional homogeneous spacetimes, we will
turn to higher-dimensional kinematical algebras. We will not discuss
gauge theories based on these algebras. Nevertheless, we find that the
construction of metric Lie algebras out of these higher-dimensional
kinematical algebras, using the double extension bootstrap, will lead
to interesting Lie algebras, some of which appear to be novel while
others are well-known in the literature. We finally provide a version
of the cube of Bacry--Lévy-Leblond for metric Lie algebras in any
dimension. This means that we find a particular double extension of
the (A)dS algebra that allows to take a well-defined limit, both on
the level of the algebra and on the level of the invariant metric, to
a double extension of any other kinematical Lie algebra that can be
reached by a contraction. Again, all of this extended algebras have the
same homogeneous space as the original algebra.

A busy reader or somebody only interested in the final results could
consider to look at our summary in Section \ref{sec:summary} and
Figure \ref{fig:cube} and the discussion in
Section~\ref{sec:discussion}.

This paper is organized as follows. In Section
\ref{sec:three-dimens-grav} we recall the reformulation of the
Einstein--Hilbert action in its first-order form as a CS gauge theory.
We will emphasize in particular the rôle of the Klein pair and the
choice of invariant metric. Section \ref{sec:all-kinem-chern} contains
one of the main results of this work. We discuss the possible IW
contractions of the (A)dS algebra to other three-dimensional
kinematical algebras, their nontrivial central extensions, and finally
present the CS action based on the twice trivially extended (A)dS
algebra from which all other actions of the cube \ref{fig:cube} can be
obtained as a limit. In Section \ref{sec:double-extensions-1} we
define the concept of a double extension and show how the appearance
of a metric for the three-dimensional kinematical algebras of the cube
can be understood from this perspective. Section
\ref{sec:non-contracting} discusses the three-dimensional kinematical
algebras that do not arise as limits from the (A)dS algebra, which
finishes our classification of kinematical CS theories. In Section
\ref{sec:gener-high-dimens} we apply the double extension bootstrap of
Section \ref{sec:double-extensions-1} to higher dimensional
kinematical algebras. Section \ref{sec:coadj-kinem-algebr} presents
then a generalization of the cube of Bacry--Lévy-Leblond for metric
Lie algebras. Following our summary and discussion in Sections
\ref{sec:summary} and \ref{sec:discussion}, respectively, we include
in Appendix \ref{sec:conventions} our conventions concerning Lie
algebra contractions. In Appendix \ref{sec:kinbackground} we discuss
concepts regarding kinematical Lie algebras, homogeneous spacetimes,
and Cartan connections. In Appendix \ref{sec:full-acti-equat} we
display the full action constructed in Section
\ref{sec:all-kinem-chern} with its equations of motion and gauge
transformations.

Before starting, a word on terminology: We stress throughout this work
that the specification of a kinematical algebra $\fk$ without
providing also a subalgebra $\homoh$ for the Klein pair $(\fk,\homoh)$
is not enough to define the homogeneous spacetime it acts on.
Nevertheless, our use of terminology will be more loose in the sense
that we will often refer to the homogeneous spacetimes by their
kinematical algebras. The reason for this is that often only this
structure is relevant for the problem at hand. The choice of
subalgebra $\homoh$ necessary for the Klein pair will be obvious from
our choice of basis for $\fk$.

\begin{figure}[!h]
  \centering
\tdplotsetmaincoords{80}{120}
\begin{tikzpicture}[
tdplot_main_coords,
dot/.style={circle,fill},
linf/.style={thick,->,blue},
cinf/.style={thick,->,red},
tinf/.style={thick,->},
stinf/.style={thick,->,gray},
shrink/.style={thick, ->,gray},
scale=0.35
]


\node (ads) at (0,10,10) [label=center:{\hyperref[tab:kin]{\cmark}}] {};
\node (p) at (10,10,10) [label=center:{\hyperref[tab:kin]{\cmark}}] {};
\node (car) at (10,0,10) [label=center:{\hyperref[tab:kin]{\cmark}}] {};
\node (pp) at (0,0,10) [label=center:{\hyperref[tab:kin]{\cmark}}] {};
\node (nh) at (0,10,0) [label=center:{\hyperref[tab:kin]{\xmark}}] {};
\node (g) at (10,10,0) [label=center:{\hyperref[tab:kin]{\xmark}}] {};
\node (pg) at (0,0,0) [label=center:{\hyperref[tab:kin]{\xmark}}] {};
\node (st) at (10,0,0) [label=center:{\hyperref[tab:kin]{\xmark}}] {};

\draw[shorten >=0.2cm,shorten <=0.2cm, linf] (ads) -- node [sloped,below]{}(p);

\draw[shorten >=0.2cm,shorten <=0.2cm,linf] (nh) -- (g);
\draw[shorten >=0.2cm,shorten <=0.2cm,linf] (pp) -- (car);
\draw[shorten >=0.2cm,shorten <=0.2cm,linf,dashed] (pg) -- (st);

\draw[shorten >=0.2cm,shorten <=0.2cm,cinf,double] (ads) -- node [sloped,above] {} (nh);
\draw[shorten >=0.2cm,shorten <=0.2cm,cinf,double,dashed] (pp) -- (pg);
\draw[shorten >=0.2cm,shorten <=0.2cm,cinf] (p) -- (g);
\draw[shorten >=0.2cm,shorten <=0.2cm,cinf] (car) -- (st);

\draw[shorten >=0.2cm,shorten <=0.2cm,tinf,double] (ads) -- node [sloped,above] {} (pp);
\draw[shorten >=0.2cm,shorten <=0.2cm,tinf] (p) -- (car);
\draw[shorten >=0.2cm,shorten <=0.2cm,tinf] (g) -- (st);
\draw[shorten >=0.2cm,shorten <=0.2cm,tinf,double,dashed] (nh) -- (pg);


\node (adsu1) at (-\shi ,10+\shi,10+\shi) [label=center:{\hyperref[tab:mostgen]{\cmark}}, label=above:(Anti-)de Sitter] {};
\node (pu1) at (10+\shi,10+\shi,10+\shi) [label=center:{\hyperref[tab:mostgen]{\cmark}}, label=above:Poincaré] {};
\node (nhu1label)  at (-\shi,10+\shi,-\shi)  [label=center:{\hyperref[tab:mostgen]{\cmark}},label=below:(A)dS-Galilei] {};
\node (nhu1) at (-\shi,10+\shi,-\shi) [label=center:{\hyperref[tab:mostgen]{\cmark}}] {};
\node (ppu1) at (-\shi,-\shi,10+\shi) [label=center:{\hyperref[tab:mostgen]{\cmark}}, label=above:(A)dS-Carroll] {};
\node (gu1) at (10+\shi,10+\shi,-\shi) [label=center:{\hyperref[tab:mostgen]{\cmark}}, label=below:Galilei] {};
\node (pgu1) at (-\shi,-\shi,-\shi) [label=center:{\hyperref[tab:mostgen]{\cmark}}, label=below:Para-Galilei] {};
\node (caru1) at (10+\shi,-\shi,10+\shi) [label=center:{\hyperref[tab:mostgen]{\cmark}}, label=above:Carroll] {};
\node (stu1) at (10+\shi,-\shi,-\shi) [label=center:{\hyperref[tab:mostgen]{\cmark}}, label=below:Static] {};

\draw[linf] (adsu1) -- node [sloped,below]{$0 \leftarrow \Lambda$}  (pu1);

\draw[shorten >=0.2cm,shorten <=0.2cm,linf] (nhu1) -- (gu1);
\draw[shorten >=0.2cm,shorten <=0.2cm,linf] (ppu1) -- (caru1);
\draw[shorten >=0.2cm,shorten <=0.2cm,linf,dashed] (pgu1) -- (stu1);

\draw[shorten >=0.2cm,shorten <=0.2cm,cinf,double] (adsu1) -- node [sloped,above] {$\ec=\frac{1}{c} \to 0$} (nhu1);
\draw[shorten >=0.2cm,shorten <=0.2cm,cinf,double,dashed] (ppu1) -- (pgu1);
\draw[shorten >=0.2cm,shorten <=0.2cm,cinf] (pu1) -- (gu1);
\draw[shorten >=0.2cm,shorten <=0.2cm,cinf] (caru1) -- (stu1);

\draw[shorten >=0.2cm,shorten <=0.2cm,tinf,double] (adsu1) -- node [sloped,above] {$0 \leftarrow \tau$} (ppu1);
\draw[shorten >=0.2cm,shorten <=0.2cm,tinf] (pu1) -- (caru1);
\draw[shorten >=0.2cm,shorten <=0.2cm,tinf] (gu1) -- (stu1);
\draw[shorten >=0.2cm,shorten <=0.2cm,tinf,double,dashed] (nhu1) -- (pgu1);


\draw[shorten >=0.2cm,shorten <=0.2cm,shrink] (adsu1) -- node [sloped,below] {$0 \leftarrow \alpha, \beta$}(ads) ;
\draw[shorten >=0.2cm,shorten <=0.2cm,shrink] (pu1) --(p) ;
\draw[shorten >=0.2cm,shorten <=0.2cm,shrink] (ppu1) --(pp) ;
\draw[shorten >=0.2cm,shorten <=0.2cm,shrink] (nhu1) --(nh) ;
\draw[shorten >=0.2cm,shorten <=0.2cm,shrink] (nhu1) -- (nh) ;
\draw[shorten >=0.2cm,shorten <=0.2cm,shrink] (caru1) --(car) ;
\draw[shorten >=0.2cm,shorten <=0.2cm,shrink] (pgu1) --(pg) ;
\draw[shorten >=0.2cm,shorten <=0.2cm,shrink] (gu1) --(g) ;
\draw[shorten >=0.2cm,shorten <=0.2cm,shrink] (stu1) --(st) ;

\end{tikzpicture}  
\caption{This tesseract describes the limits starting with the
  (anti-)de Sitter theories extended by two (trivial) central
  extensions. They are added such that they are nontrivial in the
  nonrelativistic ($\ec \to 0$) limit and render all theories
  well-defined (\cm) in the flat ($\el \to 0$) and ultrarelativistic
  ($\et \to 0$) limit. This spans the outer cube which is described in
  Table \ref{tab:mostgen}. The $\alpha,\beta \to 0$ limit trivializes
  the central extensions and leads to the inner cube. The
  nonrelativistic theories are, unlike the carrollian, not
  well-defined anymore. This is due to degeneracy of their invariant
  metric as indicated by a cross (\xm) and can be seen in Table
  \ref{tab:kin}. See also Table \ref{tab:CSlimit} for a summary of the
  properties of the algebras appearing in the tesseract.}
\label{fig:cube}
\end{figure}

\section{Chern--Simons theory and homogeneous spacetimes}
\label{sec:three-dimens-grav}

We briefly review the reformulation of the Einstein--Hilbert action as
a gauge theory of Chern--Simons type (for more details see
\cite{Witten:1988hc}). In Section \ref{sec:CS3dhomo} we will then
frame the discussion in terms of homogeneous spacetimes. We argue that
it is possible to write equations of motion for any three-dimensional
spatially isotropic spacetime as classified in
\cite{Bacry:1968zf,Figueroa-OFarrill:2018ilb} and explain what is
necessary to also provide an action principle.

\subsection{Three-dimensional gravity as a Chern--Simons theory}
\label{sec:3d-CS}

It is well-known that the first order form of the Einstein--Hilbert
action in three dimensions can be written as a Chern--Simons
theory~\cite{Achucarro:1987vz,Witten:1988hc}
\begin{equation}
  \label{eq:CSaction}
S_{\mathrm{CS}}[A] =\int   \langle A \wedge \dd A + \frac{1}{3} [A,A] \wedge A \rangle \, ,
\end{equation}
where a wedge product is understood in the commutator.

In the first order formulation of general relativity the fundamental
fields are the vielbein $e$ and the dualized spin-connection $\omega$,
which can be combined into the connection $A$ of the Chern--Simons
theory
\begin{align}
  \label{eq:conn}
  A
  &=e+\omega
    =e^{\mu} \hat \Pt_\mu+\omega^{\mu} \hat \Jt_\mu \,.
\end{align}
The connection $A$ is then a Lie algebra valued one form with the
Lie algebra given by
$\mathfrak{ads} = \mathfrak{so}(2,2)\simeq \mathfrak{sl}(2,\R) \oplus
\mathfrak{sl}(2,\R) \simeq \mathfrak{so}(2,1)\oplus
\mathfrak{so}(2,1)$,
$\mathfrak{ds}=\mathfrak{so}(3,1) \simeq \mathfrak{sl}(2,\C)$ (where
$\mathfrak{sl}(2,\C)$ is understood as a real Lie algebra) and
$\mathfrak{poi}=\mathfrak{iso}(2,1)$, for negative, positive and
vanishing cosmological constant, respectively. These algebras are
explicitly given by
\begin{subequations}
  \label{eq:AdSdSalgebra}
  \begin{align}
    \label{eq:AdSdSalgebra-a}
  \left[ \hJ_\mu, \hJ_\nu \right]  &=  \epsilon\indices{_{\mu\nu}^{\rho}} \, \hJ_{\rho} 
  \\
  \left[ \hJ_\mu , \hP_\nu \right]  &=  \epsilon\indices{_{\mu\nu}^{\rho}} \, \hP_{\rho}
  \\
  \left[ \hP_\mu , \hP_\nu \right] & = \pm \el^{2} \epsilon\indices{_{\mu\nu}^{\rho}} \, \hJ_{\rho}  \,.
\end{align}
\end{subequations}
The upper sign refers to $\mathfrak{ads}$, the lower sign to
$\mathfrak{ds}$ and $\mathfrak{poi}$ is given by the $\el \to 0$
limit.\footnote{%
  The above mentioned splitting of the $\mathfrak{ads}$ algebra is
  made explicit by the basis change
  $\hat{\Lt}^{\pm}_\mu=\tfrac{1}{2}\left(\hJ_\mu\pm\Lambda^{-1}
    \hP_{\mu}\right)$.} Indices are raised and lowered with a mostly plus
metric (our conventions are summarized in Appendix
\ref{sec:conventions}). In each case the Lie algebra corresponds to
the isometry algebra of the maximally symmetric spacetime of the
theory given by anti-de Sitter, de Sitter, and Minkowski space.

Having chosen a gauge algebra, the Chern--Simons theory is not
completely defined without choosing a ``trace'', which was denoted by
$\langle \quad \rangle$ in \eqref{eq:CSaction}. We will call this
symmetric, invariant, nondegenerate bilinear form on the Lie algebra
an \emph{invariant metric} where by invariance we mean that
\begin{equation}
  \Zt\cdot\langle \Xt,\Yt\rangle\equiv\langle [\Zt,\Xt],\Yt \rangle+\langle \Xt,[\Zt,\Yt] \rangle = 0
  \end{equation}
  for all Lie algebra elements $\Xt,\Yt,\Zt$.\footnote{%
    We will often abuse terminology and refer to a degenerate
    invariant symmetric bilinear form by degenerate (invariant)
    metric. If not indicated otherwise, we mean by an invariant metric
    a nondegenerate bilinear form.} For simple complex Lie algebras it
  is given up to a normalization by the Killing form. But, as
  discussed, $\mathfrak{ads}$ is semisimple as is the complexification
  of $\mathfrak{ds}$ (as a real Lie algebra $\mathfrak{ds}$ is of
  course simple). Therefore, the invariant metric of the real Lie
  algebras is parametrized by \emph{two} parameters $\mugr$ and
  $\muex$
\begin{subequations}
  \label{eq:AdSdSinvm}
\begin{align}
  \langle \hJ_\rho , \hJ_\lambda \rangle_{\muex} &= \muex \, \eta_{\rho\lambda}
  \\
  \langle \hP_\rho , \hJ_\lambda\rangle_{\mugr} &= \mugr \, \eta_{\rho\lambda}
  \\
  \langle \hP_\rho , \hP_\lambda \rangle_{\muex} &= \pm \el^{2} \, \muex\, \eta_{\rho\lambda} \,.
\end{align}
\end{subequations}
This is the most general invariant metric and it is nondegenerate for
$\mugr^{2} \mp \el^{2} \muex^{2} \neq 0$. In particular, the metric is
nondegenerate even in the $\Lambda\rightarrow 0$ limit provided
$\mu\neq 0$, even though $\mathfrak{poi}$ is not semisimple.

It is clear that this metric for $\mathfrak{poi}$ cannot come from its
Killing form, that is degenerate due to Cartan's criterion, which
shows that semisimplicity is not a necessary criterion for the
existence of an invariant metric. Rather, the $\mathfrak{poi}$ algebra
has the specific form of a \emph{double extension} which guarantees
the existence of an invariant metric
\cite{Medina1985,FigueroaO'Farrill:1995cy}. We will return to the
topic of double extensions in Section \ref{sec:double-extensions-1}.
For now let us note only that the double extension structure of the
Poincaré algebra in three dimensions does not generalize to higher
dimension. On the other hand, the invariant metric proportional to
$\muex$, coming from the Killing form of the semisimple
$\mathfrak{(a)ds}$ algebra, is generic.

Having defined a metric on the Lie algebra one can now plug the
connection \eqref{eq:conn} into the Chern--Simons action
\eqref{eq:CSaction} and use \eqref{eq:AdSdSalgebra} and
\eqref{eq:AdSdSinvm} to obtain
\begin{align}
  \label{eq:CSeo}
  S_{\mathrm{CS}}[e,\omega]
  &=
  \int \langle 2 e \wedge  R + \frac{1}{3} [e,e] \wedge e- \dd( \omega  \wedge e) \rangle_{\mugr} 
  + \int \langle e \wedge  T+ \omega  \wedge \dd \omega + \frac{1}{3} [\omega,\omega] \wedge \omega \rangle_{\muex} \nonumber \\
  &= S_{\mathrm{EH}} + S_{\mathrm{E}\chi} \, ,
\end{align}
where the dualized curvature two-form is given by
$R=\dd \omega + \frac{1}{2} [\omega,\omega]$ and the torsion two-form
by $T=\dd e +[\omega,e]$. The first term, with coupling $\mugr$, is
proportional to the standard Einstein--Hilbert action of general
relativity in the first order formulation. The second ``exotic'' term
with coupling $\muex$, has the interesting feature that it does not
change the equations of motion~\cite{Witten:1988hc}, as we will now
show explicitly
\begin{align}
 \vd S_{\mathrm{CS}} 
  &=
   \int \langle 2 F  \wedge \vd A \rangle - \int \dd \langle A  \wedge \vd A\rangle
  \\
  \nonumber 
  &= \int \langle (2   R + [e,e]) \wedge  \vd e+ 2 T  \wedge \vd \omega \rangle_{\mugr} - \int \dd \langle e \wedge  \vd \omega+  \omega \wedge  \vd e \rangle_{\mugr}  \\
  &\quad +  \int \langle 2 T \wedge  \vd e +  (2 R + [e,e]) \wedge  \vd \omega\rangle_{\muex} - \int \dd \langle e  \wedge \vd e + \omega  \wedge \vd \omega\rangle_{\muex} \,.
\end{align}
Thus, in each case the Chern--Simons flatness condition
\begin{equation}
\label{eq:CSflat}
  F=\dd A + \tfrac{1}{2} [A,A]=0
\end{equation}
  implies, for generic $\mugr$ and $\muex$, the equations of motion
\begin{align}
  R^{\lambda}&= \mp \frac{\el^{2}}{2} \epsilon\indices{^{\lambda}_{\nu\rho}} e^{\nu} \wedge e^{\rho}
  &
  T^{\lambda} &=0 \,.
\end{align}
The resulting geometries are therefore quotients of three-dimensional
Minkowski or (A)dS.

While the addition of the exotic term leaves the equations of motion
inert, let us emphasize that this change of the action has
repercussion on the thermodynamics of the solutions
\cite{Solodukhin:2005ah, Park:2006gt, Townsend:2013ela} and, consequently, might change the quantum
theory~\cite{Witten:1988hc}. We will therefore try to be as general as
possible when providing invariant metrics for other Lie algebras.

We parametrize the gauge transformations
$\vd A= \dd \varepsilon + [A,\varepsilon]$ by
$\varepsilon = \zeta^{\mu} \hP_{\mu} + \lambda^{\mu} \hJ_{\mu}$ which
leads us to
\begin{align}
  \vd e &= \dd \zeta + [\omega,\zeta] + [e, \lambda] \, , &  \vd \omega &= \dd \lambda + [\omega, \lambda] + [e, \zeta] \, .
\end{align}
The remarkable feature that makes three dimensional gravity special is
invariance of the action, up to a boundary term, under local Lorentz
transformations \emph{and} local translations. Diffeomorphisms $\xi$
are on-shell just gauge transformations given by
$\varepsilon= \xi^{\mu}A_{\mu}$~\cite{Witten:1988hc}.

Furthermore, one can construct a metric being invariant under local
Lorentz transformations $\lambda$ by
\begin{align}
  \label{eq:gmetric}
  g =  \eta_{\mu\nu} e^{\mu} \otimes e^{\nu} \, .
\end{align}
This quantity characterizes lorentzian geometries and will in general
be lost in the kinematical limits. This metric on the tangent space of
the spacetime $\cM$ should not be confused with the earlier introduced
invariant metric of the Lie algebra.

\subsection{Chern--Simons theories for three-dimensional homogeneous spacetimes}
\label{sec:CS3dhomo}

Let us discuss the reformulation of Einstein gravity presented in the
previous section from the point of view of homogeneous spacetimes; for
a summary of relevant concepts consult Appendix
\ref{sec:kinbackground}. The three algebras $\mathfrak{ads}$,
$\mathfrak{ds}$, and $\mathfrak{poi}$ are examples of a kinematical
Lie algebra $\fk$. In order to define a homogeneous spacetime
associated to a kinematical algebra $\fk$ we need in addition the
choice of a subalgebra $\homoh \subset \fk$ that determines the
subalgebra of boosts and rotations. For all of the three algebras this
subalgebra is the three-dimensional Lorentz algebra
$\mathfrak{so}(2,1)$ in equation \eqref{eq:AdSdSalgebra-a} generated
by the elements $\{\hat{\Jt}_\mu\}$. Writing down the gauge connection
$A$, the elements along the $\homoh$ direction are associated with the
spin connection $\omega$, whereas elements along the remaining
directions $\fk/\homoh$ are associated with the dreibein $e$. Choosing
an invariant metric one can then write the CS action
\eqref{eq:CSaction} for the connection A. Projecting the equation of
motion \eqref{eq:CSflat} along the $\homoh$ direction yields the
curvature equation, while projecting along the remaining generators
$\fk/\homoh$ yields the torsion of the spacetime. Since the
homogeneous spacetimes obtained from the pair
$(\fk,\mathfrak{so}(2,1))$, with $\fk$ being any of $\mathfrak{ads}$,
$\mathfrak{ds}$, or $\mathfrak{poi}$, can be equipped with a
lorentzian structure $\eta_{\mu\nu}$ we can write down the spacetime
metric $g$ according to equation \eqref{eq:gmetric}.

Framing the derivation of the last section in this way, we see that it
is possible to write down equations of motion for all of the
three-dimensional homogeneous spacetimes classified in
\cite{Figueroa-OFarrill:2018ilb}. One simply replaces the Klein pair
of, say, $(\mathfrak{ads},\mathfrak{so}(2,1))$ by the corresponding
Klein pair $(\mathfrak{k},\homoh)$. The equations of motion
\eqref{eq:CSflat} provide a theory for the gauge field $A$ of the
underlying symmetry algebra. More precisely, the connection $A$ is the
\emph{Cartan connection} for a \emph{Cartan geometry}. These
geometries describe manifolds that are locally modeled on a particular
homogeneous spacetime, in the same way \mbox{(pseudo-)riemannian}
manifolds are locally modeled on euclidean (Minkowski) spacetime (for
more details and references see Appendix \ref{sec:kinbackground}). Of
course, the equation of motion \eqref{eq:CSflat} tells us that,
on-shell, all solutions are quotients of the model homogeneous
spacetimes, i.e., the Cartan geometry is flat.

Since a generic homogeneous spacetime does not allow for a lorentzian
structure $\eta_{\mu\nu}$, equation \eqref{eq:gmetric} is replaced by
the corresponding carrollian, galilean or static structure. With the
appropriate replacements one can write down a Cartan connection and
the equations of motion $F=0$ for all higher-dimensional kinematical
algebras.

However, there is an additional nontrivial ingredient that is needed
to construct the CS action, which is the existence of an invariant
metric on the kinematical Lie algebra. In fact, in addition to the
choice of homogeneous spacetime this is the only nontrivial (local)
ingredient in order to define a CS theory.

For the (semi)simple algebras $\mathfrak{(a)ds}$ the existence of this
metric is guaranteed, while for $\mathfrak{poi}$ it is only due to its
special form as a double extension; we will return to this point in
Section \ref{sec:double-extensions-1}. In contrast, most kinematical
algebras $\mathfrak{k}$ on which the spacetimes appearing in the
classification of \cite{Figueroa-OFarrill:2018ilb} are based, do not
allow for an invariant metric. In order to obtain a CS theory for
these kinematical algebras we will have to extend these algebras,
i.e., we will have to add additional generators and corresponding
fields to the gauge connection $A$ such that the resulting algebra can
be endowed with an invariant metric. These extensions will be modeled
to have the structure of a double extension with the minimal number of
additional generators. This choice guarantees, e.g., compared to a
deformation, that the geometries do not loose their kinematical
structure and interpretation. Note that this is a choice and there
might be other ways to realize CS theories for these homogeneous
spacetimes. Nevertheless, using our approach we will be able to
construct CS theories for all three-dimensional homogeneous
spacetimes.

\section{Kinematical limits of three-dimensional gravity}
\label{sec:all-kinem-chern}

In the previous section we have reviewed the Chern--Simons formulation
of three-dimensional gravity and subsequently generalized the
discussion to include Chern--Simons theories based on any
three-dimensional homogeneous spacetime provided they admit an
invariant metric on their Lie algebra.

Kinematical spacetimes have been classified in the seminal work of
Bacry and Lévy-Leblond~\cite{Bacry:1968zf,Bacry:1986pm} with a recent
generalization presented in~\cite{Figueroa-OFarrill:2018ilb}. Among
these spacetimes there is a class that arises as a limit from (A)dS
space. In this section we focus on this family. The remaining
so-called \emph{non-contracting kinematical spacetimes} will be dealt
with in Section~\ref{sec:non-contracting}.

For interesting Chern--Simons theories that go beyond or generalize
the kinematical setup, see, e.g.,
\cite{Hofman:2014loa,Hartong:2016yrf,Bergshoeff:2016lwr,Hartong:2017bwq,Joung:2018frr,Aviles:2018jzw}.

\subsection{Kinematical limits}
\label{sec:nonext-contr}

Since the (semi)simple $\mathfrak{(a)ds}$ algebras cannot arise as
nontrivial contractions,\footnote{%
  This follows from the fact that there exists for any Lie algebra
  contraction an inverse operation given by a Lie algebra deformation.
  This implies that a Lie algebra that is the result of a nontrivial
  contraction should allow for deformations. However, semisimple Lie
  algebras, like the $\mathfrak{(a)ds}$ algebras, cannot be deformed
  (they are rigid) and can therefore not arise from a limit~(see,
  e.g.,~\cite{FialMon:DefCon,Fialowski_2005} for further details).}
they present a natural starting point in the search for limits. We
will call a limit or action \emph{well-defined} if the invariant
metric remains nondegenerate after the limit, which means that each
field has a kinetic term $\langle A \wedge \dd A \rangle$ in the CS
action.

In the following, it will be convenient to introduce a time-space
splitting of the $\mathfrak{(a)ds}$ Lie algebra
\eqref{eq:AdSdSalgebra}. We assume the following conventions for the
generators of rotations $\Jt$, dualized boosts $\Bt_a$, time
translation $\Ht$, and spatial translations $\Pt_a$
\begin{align}
  \label{eq:split21d}
\Jt &\equiv  \hJ_0  & \Bt_a &\equiv  \hJ_a  & \Ht &\equiv \hP_0  & \Pt_a &\equiv \hP_a \, ,
\end{align}
where $a=1,2$ (for a summary of our conventions see Appendix
\ref{sec:conv-2+1-dimens}). The subalgebra $\homoh$ of
the kinematical algebra is always assumed to be generated by the set
$\{\Jt, \Bt_a\}$.

Besides the already introduced flat or vanishing cosmological constant
limit parametrized by $\Lambda \to 0$ we have two further independent
options to introduce contraction parameters when we restrict to
kinematical spacetimes~\cite{Bacry:1968zf} (which is also true under
less restrictive conditions~\cite{Figueroa-OFarrill:2018ilb}):
\begin{itemize}
\item The nonrelativistic or galilean limit corresponds intuitively to
  the limit where the light cone opens up and is parametrized by
  sending the speed of light to infinity. We will introduce a
  contraction parameter that is the inverse of the speed of light
  $c_{i} \equiv \frac{1}{c}$ so that we can send all the limits
  uniformly to zero. One ends up with (A)dS galilean spacetimes, also
  known as Newton--Hooke spacetimes or expanding/oscillating
  universes~\cite{Bacry:1968zf}. For the case of vanishing cosmological
  constant, the $c_i\rightarrow 0$ limit leads to the galilean
  spacetime.

\item The ultrarelativistic or carrollian limit corresponds to the
  limit where the light cone closes. It is an independent and
  inequivalent contraction parametrized by $\et$ which we also send to
  zero in the limit. This leads to the (A)dS-carrollian spacetimes,
  which were called para-Poincaré and inhomogeneous $SO(4)$ in
  \cite{Bacry:1968zf}. Taking both $\Lambda \rightarrow 0$,
  $\tau\rightarrow 0$ limits leads to the carrollian spacetime. These
  carrollian spacetimes are null surfaces of (A)dS and Minkowski
  space~\cite{Figueroa-OFarrill:2018ilb} (see also
  \cite{Morand:2018tke}).

  Notice that seen as an abstract Lie algebra the AdS-Carroll algebra
  is isomorphic to the Poincaré algebra. However, the associated Klein
  pairs, and consequently the associated homogeneous spacetimes,
  differ due to a different choice of subalgebra $\homoh$ for the same
  kinematical algebra. This example emphasizes the physical importance
  of considering Klein pairs instead of kinematical algebras only.

\item Keeping the cosmological constant but sending
  $\tau,\ec\rightarrow 0$ (in any order) leads to the so-called
  para-Galilei algebra. As in the above case, the para-Galilei algebra
  is isomorphic to the Galilei algebra. The homogeneous spacetimes
  associated to the two algebras are different due to a different
  choice of subalgebra $\homoh$. We want to mention the following
  subtlety: Since the boosts $\Bt_a$ in the para-Galilei algebra do
  not act effectively on $\Pt_a,\Ht$, they can be quotiented out of
  the algebra (cf.\ Appendix \ref{sec:kinbackground}). Consequently,
  the homogeneous spacetime of the para-Galilei algebra is the same
  static aristotelian spacetime as for the static algebra, cf.\
  below.\footnote{%
    For the CS theories this question of equivalence is more subtle
    and the reason we keep the para-galilean theory.}
    
\item Sending all contraction parameters to zero
  $\Lambda,\tau,c_i\rightarrow 0$ leads to the static algebra with all
  brackets identically zero, except for the commutators of angular
  momentum with boosts and spatial momentum. The static algebra
  belongs to the class of aristotelian kinematical algebras, which we
  will analyze in more detail in Section \ref{sec:aristotelian}.
\end{itemize}
These limits can be implemented in the following way. Starting from
the generators \eqref{eq:split21d} that are appropriate for the
relativistic algebras we will change to a new basis by a
three-parameter $(\ec,\et,\el)$ family of linear transformations
$g_{\ec,\et,\el}$
\begin{align}
  \label{eq:contr}
  g_{{\tiny \el,\et,\ec}}\Jt =\Jt \qquad
 g_{{\tiny \el,\et,\ec}}\Bt_{a}=(\et\ec)^{-1}\Bt_{a}\qquad
  g_{{\tiny \el,\et,\ec}}\Pt_{a} = (\el \ec)^{-1}\Pt_{a}\qquad
 g_{{\tiny \el,\et,\ec}}\Ht =(\el \et)^{-1}\Ht \, ,
\end{align}
simultaneously transforming the metric \eqref{eq:AdSdSinvm}
\begin{align}
  \label{eq:contrme}
    g_{{\tiny \el,\et,\ec}}\mugr=(\el\et)^{-1} \mugr  \qquad g_{{\tiny \el,\et,\ec}}\muex= \muex \, ,
\end{align}
where we have for completeness also added the cosmological constant
that was already used in Section~\ref{sec:three-dimens-grav}. The
resulting generators on the right hand side are taken to be the
generators for the contracted algebras. The replacements of algebra
\eqref{eq:AdSdSalgebra} and invariant metric \eqref{eq:AdSdSinvm} lead
to Table \ref{tab:kin}. The invariant metric is nondegenerate for
\begin{align}
  \label{eq:degcriterion1}
\ec( \mugr^{2}   \mp \el^{2}\et^{2} \muex^{2}) \neq 0 \,.
\end{align}
\begin{table}[h!]
  \centering
$
\begin{array}{l r r r r r}
\toprule %
                                             & \mathfrak{(a)ds}_{\parmp}              \\ \midrule
  \left[ \Jt , \Bt_{a} \right]      & \epsilon_{am}  \Bt_{m}                 \\
  \left[ \Jt , \Pt_{a} \right]      & \epsilon_{am}  \Pt_{m}                 \\ 
  \left[ \Bt_{a} , \Bt_{b} \right] & - \epsilon_{ab}  \et^{2} \ec^{2} \Jt   \\
  \left[ \Bt_{a} , \Ht \right]      & -\epsilon_{am} \et^{2} \Pt_{m}         \\
  \left[ \Bt_{a} , \Pt_{b} \right]  & -\epsilon_{ab} \ec^{2} \Ht             \\
  \left[ \Ht , \Pt_{a} \right]     & \quad \pm \epsilon_{am} \el^{2} \Bt_{m}      \\
  \left[ \Pt_{a} , \Pt_{b} \right] & \mp  \epsilon_{ab} \el^{2} \ec^{2} \Jt \\ \bottomrule
\end{array}
$
\quad\quad\quad\quad\quad
  $
\begin{array}{l r r r}
  \toprule%
                                      & \mathfrak{(a)ds}_{\parmp}                                             \\ \midrule
  \langle \Ht , \Jt \rangle           & -\mugr                                                                \\
    \langle \Pt_{a} , \Bt_{b} \rangle & \ec^{2} \mugr \delta_{ab}               \\
  \langle \Jt , \Jt \rangle           & -\muex                                                                \\
    \langle \Bt_{a} , \Bt_{b} \rangle & \ec^{2}\et^{2}\muex   \delta_{ab}              \\
  \langle \Ht , \Ht \rangle           & \mp \el^{2} \et^{2}\muex  \\
  \langle \Pt_{a} , \Pt_{b} \rangle   & \quad \pm \ec^{2} \el^{2} \muex \delta_{ab}          \\ \bottomrule
\end{array}
$
\caption{The (anti-)de Sitter spacetimes (left) and their most general
  invariant metric (right). The upper signs correspond to
  $\mathfrak{ads}$ and the lower to $\mathfrak{ds}$. The invariant
  metric remains nondegenerate when
  $\ec( \mugr^{2} \mp \el^{2}\et^{2} \muex^{2})\neq 0$.}
\label{tab:kin}
\end{table}

From the degeneracy criterion \eqref{eq:degcriterion1} we see that the
invariant metric is non-singular in the limit $\tau\rightarrow 0$ if
$\mu\neq 0$. Since also all Lie brackets are well-defined in this
limit, this means that the ultrarelativistic limit $\et \to 0$ of the
Chern--Simons action \eqref{eq:CSaction} is well-defined. This has
first been shown for the $\Lambda=0$ case in
\cite{Bergshoeff:2016soe,Bergshoeff:2017btm}, to the best of our
knowledge.

On the other hand, in the nonrelativistic limit the bilinear form is
degenerate regardless of the choice of $\mu$ and $\chi$; the treatment
of this degeneracy is the subject of Section
\ref{sec:double-extend-kinem}. The different limits relating the
kinematic theories presented in this section are summarized in the
inner cube of the tesseract of Figure \ref{fig:cube}. The check marks
on the upper face of the inner cube signify that the limiting
procedure leads to a nondegenerate invariant metric and thus to a
well-defined theory.

A comment concerning the contractions and invariant metrics is in
order. It might happen that a contraction leads to a degenerate
bilinear form even though the contracted Lie algebra permits an
invariant metric. A straightforward example is the
contraction of an arbitrary algebra with invariant metric to an
abelian one; a contraction that is always available. Any abelian
algebra permits an invariant metric, still the contracted invariant
metric might be degenerate.

The situation for the case at hand is that, using the contractions of
Table \ref{tab:kin}, we arrive at the most general invariant metric
for the lorentzian and carrollian theories. For the Galilei,
(A)dS-Galilei, para-Galilei and static case we have the freedom to add
$\langle \Ht , \Ht \rangle = \chi_{\Ht}$, but the invariant metric is
still degenerate.

\subsection{Central extensions of kinematical algebras}
\label{sec:central-extensions}

We saw in the last subsection that the nonrelativistic theories do not
inherit a nondegenerate metric in the limit $\ec\rightarrow 0$.
However, this should not come as a surprise as it is straightforward
to show that none of these algebras permits an invariant metric.

In order to see this it is enough to realize that the center
$Z(\mathfrak{g})$ of any Lie algebra $\mathfrak{g}$ that allows for an
invariant metric is orthogonal to the first derived ideal
$[\mathfrak{g},\mathfrak{g}]$. This is implied by the following chain
of equivalences
\begin{align}
  \label{eq:5}
  [Z(\mathfrak{g}), \mathfrak{g}]&= 0
  &\iff&&
  \langle [Z(\mathfrak{g}), \mathfrak{g}], \mathfrak{g} \rangle &= 0
  &\iff&&
 \langle Z(\mathfrak{g}) , [\mathfrak{g}, \mathfrak{g}] \rangle &= 0 \,,
\end{align}
where the first arrow follows from nondegeneracy of the invariant
metric and the second from invariance. The last equation gives us the
simple necessary criterion
\begin{equation}
  \label{eq:neccriterion}
  \textrm{dim}\,\mathfrak{g}=\textrm{dim} \,[\mathfrak{g},\mathfrak{g}]+\textrm{dim} \,Z(\mathfrak{g})
\end{equation}
for the existence of an invariant metric for a given algebra
$\mathfrak{g}$. This is trivially fulfilled for semisimple Lie
algebras, where $\mathfrak{g}=[\mathfrak{g},\mathfrak{g}]$, and
abelian Lie algebras, since $\mathfrak{g}=Z(\mathfrak{g})$.

Applying equation \eqref{eq:neccriterion} to the nonrelativistic
algebras we find that the criterion is not obeyed as all of them have
no or (as in the case of the static algebra) not enough central
elements. From this we conclude that further generators, e.g, in the
form of central elements have to be added to the nonrelativistic
algebras if we require these algebras to have a nondegenerate
invariant metric. Our aim in Section \ref{sec:double-extend-kinem}
will be to include these central extensions \emph{before} taking the
nonrelativistic limit, in such a way that after the contraction the
resulting algebras have a well-defined metric; partial results of this
contraction procedure (sometimes implicitly) can be found in
\cite{Papageorgiou:2009zc,Papageorgiou:2010ud,Bergshoeff:2016lwr,Bergshoeff:2016soe,Hartong:2017bwq,Joung:2018frr}.
Before we turn to the contraction procedure we study all possible
central extensions of the kinematical algebras and their invariant
metrics that appear in the cube of Chern--Simons theories. Central
extensions are \emph{trivial} if they can be removed by a redefinition
of the basis elements. Nontrivial central extensions of the algebra
are in one-to-one correspondence with the elements of the second
cohomology group of the Lie algebra (for details cf., e.g.,
\cite{Azcarraga:2011hqa}). We mention that (semi)simple Lie algebras
can only have trivial central extensions.

The (A)dS-Galilei algebra and its flat limit, the Galilei algebra,
allow for three central extensions~\cite{levygalgr}
\begin{align}
  \label{eq:galcentr}
  [\Jt, \Ht] &= \Zt_{\Jt \Ht}
  & [\Bt_{a}, \Pt_{b}]&= \epsilon_{ab} \Mt
  & [\Bt_{a},\Bt_{b}] &= \epsilon_{ab} \St
  & [\Pt_{a},\Pt_{b}] &= \pm \Lambda^{2} \epsilon_{ab} \St
\end{align}
of which only the ``mass'' $\Mt$ generalizes to higher dimension. The
Galilei algebra with the central extension $\Mt$ included is called
the Bargmann algebra. Its importance lies in the fact that its
corresponding Lie group is isomorphic to the symmetry group of
galilean systems as described, e.g., by the Schrödinger equation. The
three-dimensional Galilei algebra with both central charges $\Mt$,
$\St$ is sometimes called extended Bargmann algebra.

The Lie algebra central extension $[\Jt, \Ht] = \Zt_{\Jt \Ht}$ does
not exponentiate to an extension of the group~\cite{levygalgr} and is
therefore often omitted. Furthermore, from the point of view that
$\Ht$ should be a scalar under rotations this central extension seems
obscure. Let us check the necessary criterion \eqref{eq:neccriterion}
for the Galilei algebra again but now with the central extensions
\eqref{eq:galcentr} included. We find that the dimension of the first
derived ideal $\{\Bt_a,\Pt_a,\Mt,\St,\Zt_{\Jt\Ht}\}$ and the dimension
of the center $\{\St,\Zt_{\Jt\Ht},\Mt\}$ do not add up to the
dimension of the algebra, thus disallowing an invariant metric, unless
one of the central extensions is dropped. Indeed, the presence of
$\Zt_{\Jt\Ht}$ renders any invariant metric degenerate, whereas the
other two central extensions are needed for a nondegenerate metric. We
will see this in more detail in Sections \ref{sec:double-extend-kinem}
and \ref{sec:double-extend-galil}.

For para-Galilei we merely exchange $\Bt \leftrightarrow \Pt$ in
\eqref{eq:galcentr} and send $\el \to 0$, so infinitesimally it also
has three central extensions given by
\begin{align}
  \label{eq:pargalcentr}
  [\Jt, \Ht] &= \Zt_{\Jt \Ht}
  & [\Bt_{a}, \Pt_{b}]&= \epsilon_{ab} \Mt
  & [\Pt_{a},\Pt_{b}] &= \epsilon_{ab} \Zt_{\Pt} \,.
\end{align}

For the relativistic algebras, i.e., the upper face of the cube, we do
not need central extensions in order to get an invariant metric, as we
showed in the last section. However, in order to be complete we
mention the possible central extensions for these algebras as well.
The Lie algebras of the Poincaré and (A)dS-Carroll spacetimes permit,
like their higher dimensional analogs, no nontrivial central
extension. The flat Carroll spacetime permits, again only in $2+1$
dimensions, three central extensions given by
\begin{align}
  \label{eq:Carcentr}
  [\Bt_{a}, \Pt_{b}]&= \delta_{ab} \tilde \Zt_{\Bt\Pt} &  [\Bt_{a},\Bt_{b}] &= \epsilon_{ab} \Zt_{\Bt} &  [\Pt_{a},\Pt_{b}] &= \epsilon_{ab} \Zt_{\Pt} \,.
\end{align}
Adding any combination of these to the algebra renders the metric
degenerate.

Finally, the static Lie algebra has five nontrivial central extensions
(see, e.g.,~\cite{Andrzejewski:2018gmz}) given by combining all of the
above mentioned to
\begin{align}
  \label{eq:statcentr}
  [\Jt, \Ht] &= \Zt_{\Jt \Ht}  & [\Bt_{a}, \Pt_{b}]&= \epsilon_{ab} \Mt+\delta_{ab} \tilde \Zt_{\Bt\Pt}  &  [\Bt_{a},\Bt_{b}] &= \epsilon_{ab} \Zt_{\Bt}  &  [\Pt_{a},\Pt_{b}] &= \epsilon_{ab} \Zt_{\Pt} \, .
\end{align}
Applying criterion \eqref{eq:neccriterion} again, we see that it is
obeyed if only one central extension is added. In fact, the only
central extensions that allow the extended static algebra to have an
invariant metric are either $\tilde{\Zt}_{\Bt\Pt}$ or $\Mt$. We will
recover the latter of these extensions when studying the contraction
of the centrally extended (A)dS algebra which is the topic of the
following section.

\subsection{The tesseract of Chern--Simons theories}
\label{sec:double-extend-kinem}

In the previous section we have not dealt with the $\mathfrak{(a)ds}$
algebra, as it is a (semi)simple algebra with trivial second
cohomology group. Thus, all central extensions of the
$\mathfrak{(a)ds}$ algebra can be eliminated by a redefinition of the
generators. However, trivial central extension may turn nontrivial
under contractions. This will allow us to obtain a nondegenerate
metric for the nonrelativistic algebras.

We extend the $\mathfrak{(a)ds}$ algebra by two $\mathfrak{u}(1)$
algebras with generators $\St$ and $\Mt$. The Lie brackets of the
extended algebra are summarized in Table \ref{tab:mostgen} (for
details see Appendix~\ref{sec:most-gener-cent}). Let us emphasize that
it is essential to add these two central extensions, as well as to
consider them as independent elements of the vector space.
\begin{table}[h]
  \centering
$
\begin{array}{l r r r r r}
\toprule %
                                             & \mathfrak{(a)ds}_{\parmp}           \dis \,\mathfrak{u}(1)^{2}        \\ \midrule
  \left[ \Jt  , \Bt_{a} \right]      & \epsilon_{am}  \Bt_{m}                                                \\
  \left[ \Jt , \Pt_{a} \right]      & \epsilon_{am}  \Pt_{m}                                                \\ 
  \left[ \Bt_{a}  , \Bt_{b} \right] & - \epsilon_{ab}  \et^{2} (\ec^{2} \Jt -\beta \St)                 \\
  \left[ \Bt_{a} , \Ht \right]      & -\epsilon_{am} \et^{2} \Pt_{m}                                        \\
  \left[ \Bt_{a} , \Pt_{b} \right]  & -\epsilon_{ab} (\ec^{2} \Ht -\alpha  \Mt)                         \\
  \left[ \Ht  , \Pt_{a} \right]     & \pm \epsilon_{am} \el^{2} \Bt_{m}                                     \\
  \left[ \Pt_{a}  , \Pt_{b} \right] & \quad \mp  \epsilon_{ab} \el^{2} (\ec^{2} \Jt - \beta\St)               \\ \bottomrule
\end{array}
$
\quad\quad\quad\quad\quad
$
\begin{array}{l r r r}
  \toprule%
                                             & \mathfrak{(a)ds}_{\parmp} \dis \,\mathfrak{u}(1)^{2}                  \\ \midrule
  \langle \Ht , \Jt \rangle                  & -\mugr                                                                \\
  \langle \Pt_{a} , \Bt_{b} \rangle          & (\ec^{2} \mugr+ \alpha \beta \mu_{\Mt \St}) \delta_{ab}               \\
  \langle \Jt , \Jt \rangle                  & -\muex                                                                \\
  \langle \Bt_{a} , \Bt_{b} \rangle          & \et^{2}(\ec^{2} \muex+ \beta^{2} \mu_{\St}) \delta_{ab}              \\
  \langle \Ht , \Ht \rangle                  & \mp \el^{2} \et^{2}\muex  +\frac{1}{\ec^{2}}(\alpha^{2} \mu_{\Mt} \mp \el^{2} \et^{2}\beta^{2} \mu_{\St}) \\
  \langle \Pt_{a} , \Pt_{b} \rangle          & \pm  \el^{2} (\ec^{2}\muex+ \beta^{2} \mu_{\St}) \delta_{ab}          \\ \midrule
  \langle \Mt, \Mt \rangle           & \ec^{2} \mu_{\Mt}                                                     \\
  \langle \St,\St \rangle            & \ec^{2} \mu_{\St}                                                     \\
  \langle \Mt,\St \rangle            & \ec^{2} \mu_{\Mt\St}                                                  \\
  \langle \Jt ,\Mt \rangle               & \beta \mu_{\Mt\St}                                                    \\
  \langle \Jt ,\St \rangle               & \beta \mu_{\St}                                                       \\
  \langle \Ht ,\Mt \rangle               & \alpha \mu_{\Mt}                                                      \\
  \langle \Ht ,\St \rangle               & \alpha \mu_{\Mt\St}                                                   \\\bottomrule
\end{array}
$

\caption{Rescaled and trivially centrally extended $\mathfrak{(a)ds}$
  algebra (left) and invariant metric (right). For a well-defined
  nonrelativistic limit of the invariant metric we demand that
  $\alpha=\beta$ and $\mu_{\Mt}=\pm\Lambda^{2} \tau^{2} \mu_{\St}$.
  Under these conditions the limits to all theories are well-defined
  and lead to the outer cube of the tesseract of Figure
  \ref{fig:cube}. The upper sign represents AdS, the
  lower dS.}
\label{tab:mostgen}
\end{table}

Notice that in the limit $\alpha,\beta \to 0$ which ``turns off'' the
central extensions the brackets and metric of Table \ref{tab:kin} are
reproduced. Furthermore, the central extensions generated by $\Mt$ and
$\St$ are trivial so that the algebra displayed in Table
\ref{tab:mostgen} is related to the previous one by a shift of
generators. Since we added two new $\mathfrak{u}(1)$ generators we are
free to choose an arbitrary nondegenerate symmetric bilinear form as
invariant metric for these generators providing us with three
parameters $\mu_{\St}$, $\mu_{\Mt}$, $\mu_{\Mt\St}$. The other
components of the metric are then fixed by the above mentioned shifts
of generators. For an explicit display of the necessary replacements
consult Appendix~\ref{sec:most-gener-cent}.

As before, the algebra is well-defined for all contractions
$\Lambda,\tau,c_i$ but in contrast to the non-extended algebra of
Table~\ref{tab:kin} the invariant metric is now nondegenerate if both
of the following conditions are fulfilled:
\begin{subequations}
  \label{eq:extnondeg}
\begin{align}
  \label{eq:determinantext}
  c_i^4\left(\mu^2\mp\Lambda^2\tau^2\chi^2\right)
  +
  \alpha^2\beta^2\mu_{\Mt\St}^2
  \mp \beta^4\Lambda^2 \tau^2 \mu_{\St}^2
  +
  2c_i^2\beta
  \left(
  \alpha\mu\mu_{\Mt\St}
  \mp\beta\Lambda^2\tau^2\chi\mu_{\St}
  \right)
  & \neq 0
  \\
  (\mu_{\St}\mu_{\Mt}-\mu_{\Mt\St}^2)&\neq 0\,.
\end{align}
\end{subequations}
Let us discuss the limits separately:
\begin{itemize}
\item For the relativistic theories with $c_i\neq 0$ already the
  non-extended metric of Table \ref{tab:kin} was nondegenerate.
  Furthermore, the central elements $\Mt,\St$ are trivial from the
  point of view of these algebras.\footnote{Although we saw before
    that the flat Carroll algebra allows for nontrivial central
    extensions the extension by $\Mt$ which is the only one surviving
    the limit $\Lambda,\tau\rightarrow 0$ is not the same as the
    extension by $\tilde{\Zt}_{\Bt\Pt}$ in equation
    \eqref{eq:Carcentr}.}

  Consistent with this we see from \eqref{eq:determinantext} that it
  is possible to turn off the central extensions by setting
  $\alpha=\beta=0$ while keeping the metric nondegenerate. This limit
  also turns off the metric elements ``coupling'' the two
  $\mathfrak{u}(1)$ generators to the remaining generators of the
  kinematical algebras. The resulting Chern--Simons theories therefore
  consists of two decoupled parts: one term associated to the
  respective relativistic kinematical algebra, the other one being two
  $\mathfrak{u}(1)$ Chern--Simons actions coupled to each other
  depending on the choice of $\mu_{\Mt\St}$.

\item The algebra obtained in the nonrelativistic limit is again the
  (A)dS-Galilei algebra for $\Lambda\neq 0$, or the Galilei algebra
  with $\Lambda=0$, with the two nontrivial central extensions denoted
  by $\Mt$ and $\St$ in \eqref{eq:galcentr}. As mentioned above,
  precisely these two central extensions are needed in order to obtain
  a nondegenerate
  metric~\cite{Papageorgiou:2009zc,Papageorgiou:2010ud}.\footnote{We
    note in passing that the AdS-Galilei algebra is decomposable, in a
    similar fashion as $\mathfrak{ads}$ decomposes into
    $\mathfrak{sl}(2)\oplus\mathfrak{sl}(2)$
    \cite{Papageorgiou:2010ud} (see also \cite{Hartong:2017bwq}).}
  Consequently, equation \eqref{eq:determinantext} shows that the
  invariant metric is now nondegenerate given that the central
  extensions are not turned off, $\alpha,\beta\neq 0$. However, due to
  our choice of parametrization the metric element
  $\langle \Ht,\Ht\rangle$ generically blows up in the nonrelativistic
  limit. In order to avoid this, we set
  \begin{align}
    \label{eq:nonwelldef}
    \alpha&=\beta & \mu_{\Mt}&=\pm\Lambda^{2} \tau^{2} \mu_{\St}
  \end{align}
  thus restricting to a four-parameter family of possible metrics. The
  invariant metric of the limit is the most general one if we add the
  term
  \begin{align}
    \label{eq:galHtHt}
    \langle \Ht , \Ht \rangle = \chi_{\Ht}
  \end{align}
  which does not arise from the limit.
\item The static algebra and its corresponding metric are obtained,
  again demanding \eqref{eq:nonwelldef}, in the limit
  $\Lambda,c_i,\tau\rightarrow 0$. The metric is nondegenerate
  provided that $\alpha, \mu_{\Mt \St} \neq 0$. From the two possible
  central extensions $\Mt$ and $\tilde{\Zt}_{\Bt\Pt}$ that allow the
  static algebra to have a nondegenerate metric the contraction
  procedure realizes the former option. The central extension $\St$
  trivializes. Notice, however, that it is not possible to decouple
  the field along the $\St$ direction while preserving a nondegenerate
  metric. The action based on the metric and algebra of Table
  \ref{tab:mostgen} in the static limit
  $\Lambda,\tau,c_i\rightarrow 0$ thus describes the aristotelian
  spacetime of the nontrivially centrally extended static algebra
  coupled to a $\mathfrak{u}(1)$ field.\footnote{This means this is
    strictly speaking not the theory with solely the $\Mt$ central
    extension we discussed in Section \ref{sec:central-extensions},
    which is due to our choice of contraction and invariant metric.}
\end{itemize}

We have thus succeeded in providing a nondegenerate metric --- and
therefore also well-defined Chern--Simons theories --- for all
theories based on kinematical spacetimes that can be obtained from the
three-dimensional (A)dS spacetimes as a limit. The situation is
summarized in the tesseract of Chern--Simons theories in Figure
\ref{fig:cube}. The inner cube of the tesseract corresponds to the
limit in which the (from the (A)dS perspective) trivial central
extensions $\Mt,\St$ are turned off. In this case, only the theories
based on the relativistic algebras are well-defined, as designated by
check marks on the respective corners. On the other hand, with the
help of the two central extensions all algebras can be equipped with
nondegenerate metrics.

\subsection{The most general action}
\label{sec:mostgenaction}

Based on the algebra and invariant metric for the trivially
centrally-extended $\mathfrak{(a)ds}$ algebra of Table
\ref{tab:mostgen} we are now able to construct the Chern--Simons
action~\eqref{eq:CSaction} that allows to obtain all other theories of
the tesseract of Figure~\ref{fig:cube} by taking various
limits.\footnote{%
  This action should be considered the most general one given the most
  general invariant metric and given our construction, which
  identifies the vielbein and the spin-connection (which we choose in
  the case of symmetric spaces to be based on the canonical connection
  of the symmetric space, see Section XI.3 in \cite{Kobayashi2})
  according to the underlying (symmetric) homogeneous spacetime. For a
  complementary perspective we refer to
  \cite{Mielke:1991nn,Baekler:1992ab,Blagojevic:2003vn,Cacciatori:2005wz}
  and the discussions in~\cite{Giacomini:2006dr}.}

The gauge field for this action
\begin{equation}
  \label{eq:3}
  A=A^{(0)}+A^{(\mathfrak{u}(1)^{2})}
\end{equation}
is given by a sum of the $\mathfrak{so}(2,2)$ or $\mathfrak{so}(3,1)$
gauge field \eqref{eq:conn} denoted by $A^{(0)}$ and gauge field
$A^{(\mathfrak{u}(1)^{2})}$ with generators $\St$ and $\Mt$. Written
explicitly we have
\begin{equation}
A=h\,\Ht + p^{a}\,\Pt_{a} + j \,\Jt+b^{a}\,\Bt_{a} + s\,\St+ m \,\Mt\,.\label{eq:A}
\end{equation}

The most general action can then be separated into three parts
\begin{equation}
  \label{eq:4}
  S=S_{\tiny\textrm{CS}}[A]=S_{\tiny\textrm{CS}}[A^{(0)}]+S_{\tiny\textrm{CS}}[A^{(\mathfrak{u}(1)^{2})}]+S^{(\alpha,\beta)}_{\tiny\textrm{I}}\,.
\end{equation}
The action $S_{\tiny\textrm{CS}}[A^{(0)}]$ contains contributions of
the fields $h,p^{a},j,b^{a}$ with the invariant metric taken to be the
one given in Table \ref{tab:mostgen} with $\alpha,\beta \to 0$. When
written in terms of $e$ and $\omega$ this is the same as
\eqref{eq:CSeo}. The second term is the abelian Chern--Simons action
based on $A^{(\mathfrak{u}(1)^{2})}=s\,\St+m \,\Mt$, explicitly given
by
\begin{equation}
  \label{eq:6}
  S_{\tiny\textrm{CS}}[A^{(\mathfrak{u}(1)^{2})}]=\int \langle A^{(\mathfrak{u}(1)^{2})}\wedge \dd A^{(\mathfrak{u}(1)^{2})}\rangle=\ec^2\int \left(\mu_{\St}\,s \wedge \dd s+\mu_{\Mt}\, m \wedge \dd m+2\mu_{\Mt\St}\, m \wedge \dd s\right)\,.
\end{equation}

The term $S^{(\alpha,\beta)}_{\tiny\textrm{I}}$ contains all the
fields together with the $\alpha,\beta$ dependent terms of the
invariant metric in Table \ref{tab:mostgen}, that induce couplings
between the $\mathfrak{u}(1)$ fields and the fields associated to the
vielbein $e$ and the spin-connection $\omega$.

Even though specifying the Lie algebra and the invariant metric is
sufficient to extract the equations of motion and gauge
transformations, we still present them fully decomposed and for
completeness in Appendix \ref{sec:full-acti-equat}.

\section{Double extensions}
\label{sec:double-extensions-1}

The fact that the Poincaré algebra in $D+1$ spacetime dimensions
$\mathfrak{iso}(D,1)$ does not have a nondegenerate invariant bilinear
metric has been one of the obstacles for attempts of constructing a
theory of gravity invariant under the full Poincaré
group.\footnote{Nevertheless, there exist higher rank invariant forms
  on the Poincaré algebra that can be used to construct Chern--Simons
  theories in dimensions higher than three, see, e.g.,
  \cite{Chamseddine:1989nu,Chamseddine:1990gk,Banados:1996hi}.} This
is related to $\mathfrak{iso}(D,1)$ not being semisimple, in contrast
to the (A)dS algebra in $D+1$ dimensions. The magic of three
dimensions, as observed by Witten in his seminal paper
\cite{Witten:1988hc}, lies in the observation that only in this case
it is possible to find a metric for $\mathfrak{iso}(2,1)$ although
this algebra is still not semisimple. As all other magic phenomena
this one too does not defy comprehension, in this particular case
coming under the mundane term \emph{double extended algebra}.

An algebra belonging to the class of double extensions comes
automatically with a nondegenerate invariant metric. In the last
sections we have seen a number of examples already in the form of the
above mentioned $\mathfrak{iso}(2,1)$, but also the Carroll algebras
and the centrally extended galilean algebras belong to this class. In
fact, any Lie algebra with a nondegenerate invariant metric that is
not a direct sum of semisimple and abelian algebras is a double
extension; we will make this statement more precise below.

In the following we will first introduce the concept of double
extensions in Section \ref{sec:double-extensions-def} and present the
structure theorem of metric Lie algebras by Medina and Revoy. In
Sections \ref{sec:nonext-lie-algebr} and \ref{sec:double-extend-galil}
we will then show how the existence of invariant metrics for the
kinematical algebras appearing in the tesseract of Chern--Simons
theories can be understood from the perspective of double extensions.

\subsection{Definition of double extensions}
\label{sec:double-extensions-def}

In our definition of double extensions we will follow
\cite{Medina1985, FigueroaO'Farrill:1995cy} but stick to a discussion
in terms of basis elements. For a basis-independent definition consult
the original works.

Suppose $(\mathfrak{g},\langle \,\, , \, \rangle_{\mathfrak{g}})$ is a
Lie algebra with nondegenerate invariant metric, i.e., we have the
relations
\begin{align}
  \label{eq:16}
  [\Gt_{i},\Gt_{j}] &=f\indices{_{ij}^{k}} \Gt_{k}
  &
    \langle \Gt_i,\Gt_j\rangle&=\Omega^{\mathfrak{g}}_{ij}
  &
    f\indices{_{ij}^{k}}\Omega^{\mathfrak{g}}_{kl}+f\indices{_{il}^{k}}\Omega^{\mathfrak{g}}_{kj}&=0\,
\end{align}
where $\Omega^{\mathfrak{g}}_{ij}$ denotes the nondegenerate symmetric
matrix of metric elements and $f\indices{_{ij}^{k}}$ the structure
constants of the algebra.

Let $\mathfrak{h}$ be a Lie algebra with generators $\Ht_\alpha$ and
structure constants $f\indices{_{\alpha\beta}^{\gamma}}$ that act on
$(\mathfrak{g},\langle \,\, , \,\rangle_{\mathfrak{g}})$ via
antisymmetric derivations, i.e., we have the relations
\begin{equation}
  \label{eq:14}
 [\Ht_{\alpha},\Gt_{i}]=f\indices{_{\alpha i}^{j}} \Gt_{j}
\end{equation}
and
\begin{align}
  \label{eq:antisymexp}
  f\indices{_{\alpha i}^{k}}\Omega_{kj}^{\mathfrak{g}}+f\indices{_{\alpha j}^{k}}\Omega_{ki}^{\mathfrak{g}}=0 \,.
\end{align}
Let $\mathfrak{h}^*$ denote the dual of $\mathfrak{h}$. Then the Lie
algebra $\mathfrak{d}=D(\mathfrak{g},\mathfrak{h})$ defined on the
vector space direct sum
$\mathfrak{g} \dot + \mathfrak{h} \dot +\mathfrak{h}^{*}$ (spanned by
$\Gt_{i}$, $\Ht_{\alpha}$ and $\Ht^{*\alpha}$, respectively) by
\begin{subequations}
\begin{align}
 [\Gt_{i},\Gt_{j}]                            & =f\indices{_{ij}^{k}} \Gt_{k}+f\indices{_{\alpha i}^{k}} \Omega_{kj}^{\mathfrak{g}} \Ht^{*\alpha}
                    \label{eq:GG}           \\
 [\Ht_{\alpha},\Gt_{i}]                       & =f\indices{_{\alpha i}^{j}} \Gt_{j}
                         \label{eq:HG}      \\
 [\Ht_{\alpha},\Ht_{\beta}]                   & =f\indices{_{\alpha \beta}^{\gamma}} \Ht_{\gamma}
                             \label{eq:HH}  \\
 [\Ht_{\alpha},\Ht^{*\beta}]                   & =-f\indices{_{\alpha \gamma}^{\beta}} \Ht^{*\gamma}
                             \label{eq:HHd} \\
  [\Ht^{*\alpha},\Gt_{j}]                      & =0
                          \label{eq:HdG}    \\
  [\Ht^{*\alpha},\Ht^{*\beta}]                  & =0
                              \label{eq:HdHd}
\end{align}
\end{subequations}
is called a \emph{double extension of $\mathfrak{g}$ by
  $\mathfrak{h}$}. It permits the invariant metric
\begin{align}
  \Omega_{ab}^{\mathfrak{d}}= \bordermatrix{~ & \Gt_{j}                    & \Ht_{\beta}                       & \Ht^{* \beta} \cr
                            \Gt_{i}           & \Omega_{ij}^{\mathfrak{g}} & 0                                 & 0  \cr
                             \Ht_{\alpha}     & 0                          & h_{\alpha\beta}                   & \delta\indices{_{\alpha}^{\beta}} \cr
                             \Ht^{ * \alpha}     & 0                          & \delta\indices{^{\alpha}_{\beta}} & 0 \cr}
\end{align}
where $h_{\alpha\beta}$ is some arbitrary (possibly degenerate or zero)
invariant symmetric bilinear form on $\mathfrak{h}$.

Thus, double extensions provide another class of Lie algebras with
nondegenerate metric, in addition to semisimple and abelian ones. In a
sense to be made more precise below, these classes are the building
blocks for all Lie algebras with nondegenerate metric.

Given two Lie algebras with invariant metric
$(\mathfrak{g}_1,\langle \,\, , \,\rangle_1)$ and
$(\mathfrak{g}_2,\langle \,\, , \,\rangle_2)$ a new Lie algebra with
invariant metric can be obtained by taking their direct sums and the
orthogonal direct product metric
\begin{equation}
  \label{eq:18}
  (\mathfrak{g}_1\oplus\mathfrak{g}_2,\langle \,\, , \, \rangle_1\dot+\langle \,\, , \,\rangle_2).
\end{equation}
A Lie algebra that can be written in this way is called
\emph{decomposable} otherwise \emph{indecomposable}. An example of a
decomposable Lie algebra is the $\mathfrak{ads}$ algebra
$\mathfrak{so}(2,2)=\mathfrak{sl}(2,\mathbb{R})\oplus\mathfrak{sl}(2,\mathbb{R})$.

The proof that \emph{all} metric Lie algebras can be obtained by
direct sums and double extensions of simple and one-dimensional Lie
algebras is due to the following structure theorem of Medina and Revoy
\cite{Medina1985} (with refinements due
to~\cite{FigueroaO'Farrill:1995cy}).
\begin{theorem}\label{thm:de}
Every indecomposable Lie algebra which permits an invariant metric
  is either:
  \begin{enumerate}
  \item A simple Lie algebra.
  \item A one-dimensional Lie algebra.
  \item A double extended Lie algebra $D(\mathfrak{g},\mathfrak{h})$ where:
    \begin{enumerate}
    \item $\mathfrak{g}$ has no factor $\mathfrak{p}$ for which
      $H^{1}(\mathfrak{p},\R) = H^{2}(\mathfrak{p},\R) = 0$.
      \label{item:MRg}
    \item $\mathfrak{h}$ is either simple or one-dimensional.
      \label{item:MRhsimp}
    \item $\mathfrak{h}$ acts on $\mathfrak{g}$ via outer derivations.
      \label{item:MRhout}
    \end{enumerate}
  \end{enumerate}
  Since every decomposable Lie algebra can be obtained from the
  indecomposable ones this theorem describes how all of them can be
  generated.
\end{theorem}

In the following we will use the double extension structure in a more
constructive way to answer the question: Given a Lie algebra without
invariant metric is it possible to add new generators, largely
preserving the commutation relations, such that the resulting new
algebra has an invariant metric? The double extension structure yields
a method to bootstrap a new algebra out of the original one. Suppose
the generators of the original algebra can be grouped in two
subalgebras $\mathfrak{g}$, $\mathfrak{h}$ with commutation relations 
\begin{equation}
  \label{eq:17}
  [\mathfrak{g},\mathfrak{g}]\subset \mathfrak{g}\qquad [\mathfrak{h},\mathfrak{g}]\subset \mathfrak{g}\qquad [\mathfrak{h},\mathfrak{h}]\subset \mathfrak{h}\,
\end{equation}
such that $\mathfrak{g}$ has a metric $\Omega^{\mathfrak{g}}$ invariant under the
action of both $\mathfrak{g}$ and $\mathfrak{h}$. Using the double
extension structure we can then construct a new algebra with
nondegenerate metric by adding new generators $\Ht^\alpha\in \fh^*$.
We will thus add $\textrm{dim}(\mathfrak{h})$ additional generators.
Notice that these additional generators will be central extensions for
the original algebra if $\mathfrak{h}$ is abelian.

In general, there will exist several choices to group the generators
of the original algebra into $\mathfrak{g}$ and $\mathfrak{h}$. A
choice that is always possible is to take $\mathfrak{g}$ the trivial
algebra and let $\mathfrak{h}$ contain all the generators. The
resulting double extension, that is twice as large as the
original algebra, is called \emph{coadjoint algebra}. However, we
will usually be interested in the case where $\mathfrak{h}$ has the
smallest dimension. In the next sections we will see this procedure at
work for the kinematical algebras of Section
\ref{sec:all-kinem-chern}.

Let us note already here that the extensions arrived at using the
procedure outlined above will be a mild extension of the original
kinematical algebra in the following sense: Although in general the
new generators $\Ht^\alpha$ are not central they always form an ideal
of the resulting extended algebra; cf.\ equations
\eqref{eq:HHd}-\eqref{eq:HdHd}. Since our definition of homogeneous
spacetime requires the action of all generators to be effective (cf.\
Appendix \ref{sec:kinbackground}), one should quotient by this ideal
when determining the Klein pair of the spacetime. The double extension
procedure therefore does not change the underlying homogeneous
spacetime but adds only additional generators, such that the metric is
well-defined. In the following, we will tacitly assume that a
corresponding number of additional fields is included in the CS
action.

The interpretation for these additional fields will depend on the
details of the model, which we will therefore leave for future work.
Let us nevertheless add some general comments. Even though, as we just
argued, our changes are rather mild, we want to emphasize that, due to
the different field content, different boundary conditions might be
required for different extensions. These change the physics, for
example in nonrelativistic setups by adding a central element related
to mass. Since boundary conditions should reflect the physical
solution space, which might differ drastically for different theories
(see, e.g, asymptotically flat, dS or AdS boundary conditions) or
applications, general statements beyond required consistency
requirements are hard to come by.

\subsection{(A)dS, Poincaré, and carrollian algebras}
\label{sec:nonext-lie-algebr}

Let us now investigate how the metric Lie algebras of
Section~\ref{sec:all-kinem-chern} that underlie the kinematical
gravitational theories can be understood as double extensions.

For the (semi)simple $\mathfrak{(a)ds}$ algebras there is not much to
say, except that they are naturally equipped with an invariant metric
given by the Killing form and can be seen as a trivial double
extension $D(\mathfrak{(a)ds},0)$. The fact that their invariant
metric is given by a two parameter family (see \eqref{eq:AdSdSinvm}
for $\Lambda\neq 0$) is due to the fact that the Lie algebra splits
over the complex numbers.

Since the Poincaré algebra in three dimensions is neither semisimple
nor abelian but permits an invariant metric the above mentioned
theorem guarantees that it is a double extension. In order to see this
from the definition of section \ref{sec:double-extensions-def}, take
$\mathfrak{g}$ to be the trivial algebra and $\mathfrak{h}$ to be
spanned by $\hJ_{\mu}$ and $\mathfrak{h}^{*}$ by $\hP^{\mu}$. The
commutation relations \eqref{eq:HH}, \eqref{eq:HHd} and
\eqref{eq:HdHd} reproduce the ones of \eqref{eq:AdSdSalgebra} with
$\Lambda=0$ when indices are raised and lowered with
$\eta_{\mu\nu}$. Thus, we have $\mathfrak{poi}=D(0,\{ \hJ_{\mu}\})$,
i.e., it is the coadjoint extension of the Lorentz algebra
$\mathfrak{so}(2,1)$. It is crucial for this to work that there are as
many rotations and boosts as there are time and spatial
translations. This, being the real magic of three dimensions, is not
the case in higher dimensions.

Since AdS-Carroll is isomorphic to the Poincaré algebra upon exchange
of $\Bt_{a}$ and $\Pt_{a}$ what we just said easily generalizes to
this case, as well as to the dS Carroll case.

For the Carroll case we take again $\mathfrak{g}$ as the trivial
algebra and consider $\mathfrak{h}$ to be generated by rotation $\Jt$
and boosts $\Bt_a$. Introducing the generators $\Pt_a=\Bt_a^*$ and
$\Ht=\Jt^*$ one reproduces the Carroll algebra. The Carroll algebra is
symmetric under the exchange of boosts and spatial momenta. Thus, one
can equally well consider it as the double extension of the trivial
algebra with $\mathfrak{h}$ generated by $\Jt$ and $\Pt$.

\subsection{Double extended (A)dS-Galilei and Galilei}
\label{sec:double-extend-galil}

Let us now turn to the nonrelativistic algebras. This will provide us
with the first nontrivial example of the double extension bootstrap
which we will treat rather explicitly.

By our recipe outlined at the end of Section
\ref{sec:double-extensions-def} we should group the generators of the
galilean algebras into two subalgebras $\mathfrak{g}$, $\mathfrak{h}$
compatible with the double extension structure. Looking at the
suggestively ordered three dimensional (A)dS Galilei algebras in
\begin{subequations}
    \label{eq:3dgal}
\begin{align}
  [\Jt,\Bt_a]&=\epsilon_{ab}\Bt_b\qquad [\Jt,\Pt_a]=\epsilon_{ab}\Pt_b\\
  [\Ht,\Bt_a]&=\epsilon_{ab}\Pt_b\qquad [\Ht,\Pt_a]=\pm\Lambda^2\epsilon_{ab}\Bt_b
\end{align}
\end{subequations}
one recognizes that $\{ \Jt, \Ht \}$ acts on $\{ \Bt_{a},\Pt_{a} \}$
as antisymmetric derivations. Since the commutators of
$\{ \Bt_{a},\Pt_{a} \}$ are abelian we can define an invariant metric
$\Omega_{kj}^{\mathfrak{g}}$ on this subalgebra and identify it as
$\mathfrak{g} \simeq \{ \Bt_{a},\Pt_{a} \}$ which implies
$f\indices{_{ij}^{k}}=0$ in \eqref{eq:GG}. The action of
$\mathfrak{h}=\{ \Jt, \Ht \}$ on $\mathfrak{g}$ fixes
$f\indices{_{\alpha i}^{j}}$ in $\eqref{eq:HG}$, and the commutation
relations in equation \eqref{eq:3dgal} show that
$f\indices{_{\alpha\beta}^\gamma}=0$ in \eqref{eq:HH}. The elements of
$\mathfrak{h}^*$ will therefore play the rôle of central extensions of
the original algebra.

In order for this choice of $\mathfrak{g}$,
$\Omega_{kj}^{\mathfrak{g}}$, and $\mathfrak{h}$ to be admissible for
a double extension one has to check that condition
\eqref{eq:antisymexp} is satisfied. Invariance under the action of
$\mathfrak{\Jt}$ restricts $\Omega_{kj}^{\mathfrak{g}}$ to
\begin{equation}
  \label{eq:19}
  \langle \Pt_a,\Pt_b\rangle=\mu_{\Pt} \delta_{ab}\qquad  \langle \Bt_a,\Bt_b\rangle=\mu_{\Bt} \delta_{ab}\qquad \langle \Bt_a,\Pt_b\rangle=\mu_{\Bt\Pt}\delta_{ab}+\tilde{\mu}_{\Bt\Pt} \epsilon_{ab}\,.
\end{equation}
From invariance under the action of $\Ht$ we find the conditions
\begin{equation}
  \label{eq:20}
  \tilde{\mu}_{\Bt\Pt}=0\,\qquad \mu_{\Pt}=\pm \Lambda^2\mu_{\Bt}\,.
\end{equation}

Note that there exist two more options for
grouping the generators into $\mathfrak{g}$ and $\mathfrak{h}$
according to \eqref{eq:17}. The choice
$\mathfrak{g}=\{\Ht,\Pt_a,\Bt_a\}$ would result in a smaller number of
additional generators, however one cannot define a nondegenerate
metric $\Omega^{\mathfrak{g}}$ on $\mathfrak{g}$, thus excluding this
case. The other choice $\mathfrak{g}=\{0\}$ corresponds to the
coadjoint extension that is possible for any algebra but leads to a
doubling of the algebra. The case $\mathfrak{g}=\{\Bt_a,\Pt_a\}$ is
therefore preferred due to the smaller number of additional generators.

Based on the two-parameter family $(\mu_{\Bt\Pt},\mu_{\Bt})$ we can
now start the double extension procedure. Since the algebra
$\mathfrak{h}$ generated by $\Ht$ and $\Jt$ is again an abelian
algebra any two-dimensional metric will be invariant. We choose the
metric to be of the form
\begin{equation}
  \label{eq:21}
  \langle \Ht,\Jt\rangle=-\mu \qquad \langle \Jt,\Jt\rangle=-\chi \qquad \langle \Ht,\Ht\rangle= \chi_{\Ht}\,.
\end{equation}
The dual elements that we denote by $\mathfrak{h}^*=\{\Jt^*,\Ht^*\}$
have the metric
\begin{equation}
  \label{eq:22}
  \langle \Jt,\Jt^*\rangle=1 \qquad \langle \Ht,\Ht^*\rangle=1\,.
\end{equation}
The double extension structure now dictates via equation \eqref{eq:GG}
the following change in the commutation relations of $\mathfrak{g}$:
\begin{subequations}
    \label{eq:38}
\begin{align}
  [\Bt_a,\Pt_b]&=\epsilon_{ab}(\mu_{\Bt\Pt}\Jt^*\pm \Lambda^2\mu_{\Bt} \Ht^*) \qquad [\Bt_a,\Bt_b]=\epsilon_{ab}(\mu_{\Bt}\Jt^*+\mu_{\Bt\Pt}\Ht^*) \\
  [\Pt_a,\Pt_b]&=\pm\Lambda^2\epsilon_{ab}(\mu_{\Bt}\Jt^*+\mu_{\Bt\Pt}\Ht^*)\,.
\end{align}
\end{subequations}
From comparison with the metric of the galilean algebras that we
obtained in Section \ref{sec:double-extend-kinem} we find that we
should identify
\begin{equation}
  \label{eq:29}
  \mu_{\Bt\Pt}=\alpha^2\mu_{\Mt\St}\qquad \quad \mu_{\Bt}=\alpha^2\mu_{\St}\,.
\end{equation}
Setting now
\begin{equation}
  \label{eq:39}
  \alpha\Mt= \mu_{\Bt\Pt}\Jt^*\pm \Lambda^2 \mu_\Bt\Ht^*\, \qquad \quad \alpha \St= \mu_{\Bt}\Jt^*+\mu_{\Bt\Pt}\Ht^*
\end{equation}
in \eqref{eq:38} and \eqref{eq:22}, we reproduce together with
\eqref{eq:3dgal} and \eqref{eq:21} precisely the commutation relations
and metric elements in Table \ref{tab:mostgen} of the galilean
theories. The freedom in choosing $\chi_{\Ht}$ in \eqref{eq:21}
corresponds to the term in \eqref{eq:galHtHt} that is an additional
freedom in the choice of invariant bilinear form for the galilean
theories not apparent from Table \ref{tab:mostgen}.

\section{Non-contracting theories}
\label{sec:non-contracting}

In the last section we were concerned with kinematical Lie algebras
that arise as contractions of the simple $\mathfrak{(a)ds}$ algebra.
These kinematical algebras are well-known since the seminal work of
Bacry and Lévy-Leblond~\cite{Bacry:1968zf}. However, there exist
kinematical algebras that do not arise as a limit but nevertheless
lead to well-defined kinematical spacetimes. These have been
classified only recently~\cite{Figueroa-OFarrill:2018ilb}, and some of
them are based on kinematical Lie algebras that are special to $2+1$
dimensions~\cite{Andrzejewski:2018gmz}. In this section we want to
study if it is possible to construct Chern--Simons theories based on
these algebras, i.e., if these algebras, or any of their possible
central extensions, allow for a nondegenerate invariant metric. If the
centrally extended algebras do not allow for an invariant metric, we
will examine whether there exists a double extension based on these
algebras with an invariant metric by construction. In case double
extensions with fewer additional generators do not exist the coadjoint
extension is always possible. Applying these three tools to construct
metric extensions of a given algebra we analyze the remaining
three-dimensional kinematical algebras in the following.

\subsection{Torsional galilean theories}
\label{sec:tors-galil-theor}

As our first class of non-contracting kinematic Lie algebras we treat
the so-called \emph{torsional galilean theories}. All of them share
the commutators
\begin{align}
\label{eq:kinbrackets}
  [\Jt,\Bt_a] & =\epsilon_{ab}\Bt_b
  & [\Jt,\Pt_a] &=\epsilon_{ab}\Pt_b
  & [\Ht,\Bt_a]&=\epsilon_{ab}\Pt_b \,.
\end{align}
For the \emph{torsional galilean-dS algebra} we add
\begin{equation}
  \label{eq:11}
  [\Ht,\Pt_a]=\gamma \epsilon_{ab}\Bt_b+(1+\gamma)\Pt_a \, ,
\end{equation}
while for the \emph{torsional galilean-AdS algebra} we add
\begin{equation}
  \label{eq:12}
  [\Ht,\Pt_a]=(1+\chi^2) \epsilon_{ab}\Bt_b+2\chi\Pt_a \,.
\end{equation}
The free parameters are constrained by $\gamma\in[-1,1),\chi>0$. For
$\chi=0$, the interpolating algebra reduces to galilean AdS, while
$\gamma=-1$ corresponds to galilean dS. Let us emphasize that the
spacetimes associated to these algebras differ for different $\gamma$
and $\chi$ which means that these parameters should not be understood
as contraction parameters. The algebras \eqref{eq:11} and
\eqref{eq:12} exhibit the same invariant structure common to all
galilean theories given by invariant one-form and invariant co-metric.
Different from the usual (A)dS-galilean or flat galilean algebras, the
invariant connection associated to the homogeneous spacetimes of
\eqref{eq:11} and \eqref{eq:12} exhibits torsion, hence their name
\cite{Figueroa-OFarrill:2018ilb}.

While these torsional galilean algebras exist in any dimension, in
$2+1$ dimensions one has an additional two-parameter family
interpolating between the two of them, labelled \emph{S12} in the
classification of \cite{Figueroa-OFarrill:2018ilb}, which is given
by \eqref{eq:kinbrackets} and
\begin{equation}
  \label{eq:13}
  [\Ht,\Pt_a]=(1+\gamma)\Pt_a-\chi \epsilon_{ab}\Pt_b+\gamma \epsilon_{ab}\Bt_{b}+\chi \Bt_{a}\,,
\end{equation}
where the parameters $\gamma,\chi$ are restricted to the same range as
above.

All torsional galilean Lie algebras have a unique central extension
given by
\begin{equation}\label{eq:centrtor}
  [\Jt, \Ht]=\Zt .
\end{equation}

Applying the necessary criterion of \eqref{eq:neccriterion} to the
torsional galilean algebras we see that none of them comes equipped
with a nondegenerate invariant metric, regardless of the presence or
absence of the central charge. Indeed, they all share, irrespective of
the presence of the central extension \eqref{eq:centrtor}, the same
most general degenerate invariant metric given by
\begin{align}
  \label{eq:invgal}
  \langle \Jt , \Jt \rangle &= \chi_{\Jt} &   \langle \Jt , \Ht \rangle &= \chi_{\Jt \Ht} &   \langle \Ht , \Ht \rangle &= \chi_{\Ht} \,.
\end{align}
Thus, it is not possible to have well-defined Chern--Simons theories
based on the above (centrally extended) algebras.

Let us therefore turn to double extensions. We will proceed as we did
for the galilean algebras in Section \ref{sec:double-extend-galil} by
splitting the algebra into subalgebras $\mathfrak{g}$ and
$\mathfrak{h}$, adding new generators corresponding to the dual of
$\mathfrak{h}$. In particular, we want to find the double extension
with the smallest number of additional generators.

We treat the cases of torsional galilean (A)dS algebra and the
interpolating algebra at once by writing the second commutator of
equations \eqref{eq:11}, \eqref{eq:12}, and \eqref{eq:13},
respectively, as
\begin{equation}
    [\Ht,\Pt_a]=\alpha \epsilon_{ab}\Bt_{b}+\beta\Pt_a+\tilde{\beta} \epsilon_{ab}\Pt_b+\tilde{\alpha} \Bt_{a},
\end{equation}
with constant $\alpha,\beta,\tilde{\alpha},\tilde{\beta}$ according to
the respective algebra.

In order to bootstrap a new double extended algebra out of this we
have to choose subalgebras $\mathfrak{g}$ and $\mathfrak{h}$. The only
choice, apart from the coadjoint extension, compatible with the double
extension structure is $\mathfrak{g}=\{\Gt_a,\Pt_a\}$ and
$\mathfrak{h}=\{\Jt,\Ht\}$.
However, this choice of $\mathfrak{g}$ and $\mathfrak{h}$ is not
compatible with a nondegenerate metric. Since $\mathfrak{h}$ is
abelian, the new generators appearing through $\mathfrak{h}^*$ are
necessarily central. Yet, we discussed already in Section
\ref{sec:tors-galil-theor} that adding central extensions is not
enough to get a nondegenerate metric.
Thus, the only option left is to take $\mathfrak{h}$ to contain all
the generators so that we end up with the coadjoint algebra of the
torsional galilean theories. More explicitly, we introduce the
elements $\{\Ht^*,\Jt^*,\Bt_a^*,\Pt_a^*\}$ together with the brackets
summarized on the left hand side of Table \ref{tab:3dtorgal}, which
leads to the coadjoint metric on the right hand side.
\begin{table}[h]
  \centering
$
\begin{array}{l r r}
\toprule %
                        & \textrm{torsional galilean}\\ \midrule
   \left[\Jt,\Xt_a^*\right] &  \epsilon_{ab}\Xt_b^*\\
  \left[\Ht,\Bt_a^*\right] &  -\tilde{\alpha}\Pt_a^*+\alpha\epsilon_{ab}\Pt_b^*\\
  \left[\Ht,\Pt_a^*\right] &  \epsilon_{ab}\Bt_b^*-\beta \Pt_a^*+\tilde{\beta}\epsilon_{ab}\Pt_b^*\\
  \left[\Bt_a,\Bt_b^*\right] &  \epsilon_{ab}\Jt^*\\
  \left[\Pt_a,\Pt_b^*\right] & \quad \epsilon_{ab}\Jt^*+\beta\delta_{ab}\Ht^*+\tilde{\beta}\epsilon_{ab}\Ht^*\\
  \left[\Pt_a,\Bt_b^*\right] &  \alpha\epsilon_{ab}\Ht^*+\tilde{\alpha}\delta_{ab}\Ht^* \\ \bottomrule
\end{array}
$
\quad\quad\quad\quad\quad
$
\begin{array}{l  r}
\toprule %
 \multicolumn{2}{r}{\text{torsional galilean}} \\ \midrule
  \langle \Jt , \Jt^* \rangle &  \chi \\
  \langle \Ht,\Ht^* \rangle &  \chi \\
  \langle \Bt_a,\Bt_b^*\rangle &  \chi \delta_{ab}\\
  \langle \Pt_a,\Pt_b^*\rangle & \chi \delta_{ab}\\
  \langle \Jt,\Jt\rangle &  \mu_{\Jt}\\
  \langle \Jt,\Ht\rangle &  \mu_{\Jt}\\
  \langle \Ht,\Ht \rangle &  \mu_{\Ht}
  \\  \bottomrule
\end{array}
$
\caption{Coadjoint extension of the torsional galilean algebras. $\Xt_a^*$ stands for $\Bt_a^*,\Pt_a^*$. The metric is nondegenerate for $\chi\neq 0$.}
\label{tab:3dtorgal}
\end{table}

\subsection{Carrollian light cone}
\label{sec:carr-light-cone}

Among the non-contracting kinematical Lie algebras the
\emph{carrollian light cone algebra} appears to be the most elusive
one. Its commutation relations are given, in addition to the usual
brackets $[\Jt,\Bt_a] =\epsilon_{ab}\Bt_b$ and
$[\Jt,\Pt_a] =\epsilon_{ab}\Pt_b$, by
\begin{align}
  \label{eq:CarrollLC}
  [\Bt_a,\Pt_b]&=-\Ht \epsilon_{ab}-\Jt \delta_{ab} &  [\Ht,\Pt_a]&=-\Pt_a &  [\Ht,\Bt_a]&=\Bt_a\, .
\end{align}
Although not obvious from this choice of basis, the carrollian light
cone algebra is a simple Lie algebra. In three dimensions it is
isomorphic to $\mathfrak{so}(3,1)$. We conclude that all central
extensions are necessarily trivial and that a nondegenerate invariant
bilinear form for the carrollian light cone is provided by the Killing
form. More explicitly, the most general invariant metric is
\begin{equation}
  \label{eq:LCinvme}
  \langle \Jt,\Jt\rangle=-\langle\Ht,\Ht\rangle=\chi \qquad \langle \Jt,\Ht\rangle=\mu\qquad \langle \Bt_a,\Pt_b\rangle=-\mu\, \delta_{ab}+\chi\,\epsilon_{ab}\,,
\end{equation}
which becomes degenerate for $\chi=\mu=0$. The two-parameter family of
metrics is again due to the fact that $\mathfrak{so}(3,1)$ splits as a
complex Lie algebra. From the fact that both the carrollian light cone
algebra and the $\mathfrak{ds}$ algebra are isomorphic to
$\mathfrak{so}(3,1)$ one should not conclude that both lead to
identical gravitational Chern--Simons theories. The point of view that
we aim to stress throughout this work is precisely that the
specification of an abstract algebra is not enough to define the
gravitational Chern--Simons theory. In particular, note that the
subalgebra $\homoh=\{\Jt,\Bt_a\}$ is different in both cases.

A puzzling feature of the Lie algebra \eqref{eq:CarrollLC} is the fact
that it is not reductive. This means that it does not have a
decomposition as $\mathfrak{g}=\homoh\oplus\mathfrak{m}$ such that
both subspaces transform as adjoint $\homoh$-modules. Acting with a
boost thus mixes up the separation of the different subspaces. Some
aspects of a Chern--Simons theory based on this kinematical algebra
are discussed in \cite{Gryb:2012qt}. We will have further comments on
potential applications in the discussion.

\subsection{Aristotelian}
\label{sec:aristotelian}

The remaining cases of kinematical algebras that do not derive from a
contraction of AdS belong to the class of aristotelian algebras. An
example for members of this class, that we saw earlier, is given by
the static algebra (once it is quotiented by the boosts). These
aristotelian algebras and their associated spacetimes have been
classified only recently \cite[Table 2]{Figueroa-OFarrill:2018ilb} and
have therefore evaded a systematic study of their metric properties.
They arise due to the fact that the action of the boosts on the
spacetime is not effective, which makes it natural to quotient by
them. From this point of view, the aforementioned para-Galilei
spacetime is then equivalent to the aristotelian static
spacetime.\footnote{%
  This construction does not necessarily generalize once the invariant
  metric is also taken into account, as we saw, e.g., for the static
  theory in Section \ref{sec:double-extend-kinem}. To distinguish
  theories based on the static spacetime with boosts from the ones
  without, we have added the specification ``aristotelian'' to the
  latter.} Let us also note that all aristotelian spacetimes contract
to the aristotelian static one.

All aristotelian Lie algebras are spanned by $\Jt$, $\Ht$ and
$\Pt_{a}$ with only
\begin{align}
\label{eq:aristo}
  [\Jt, \Pt_{a}] = \epsilon_{ab} \Pt_{b}
\end{align}
remaining of the usual kinematic brackets. Before we examine the
various options for the remaining commutators let us emphasize that in
a CS interpretation the field $\omega$ in the connection $A$ does only
have a component along the $\Jt$ direction.

\subsubsection{Aristotelian static and torsional static}
\label{sec:static}

For the aristotelian static spacetime the only nonzero commutator is
given by \eqref{eq:aristo}. For the torsional static spacetime we need
to add the commutator
\begin{align}
  [ \Ht , \Pt_{a}] = \Pt_{a} \,.
\end{align}

Both permit no nondegenerate invariant metric, as can be checked by
criterion \eqref{eq:neccriterion}. Their most general symmetric,
invariant bilinear form is again given by \eqref{eq:invgal}.

The unique central extension for the torsional static spacetime is
\begin{equation}
  \label{eq:torstcent}
 [\Jt, \Ht] = \Zt_{\Jt \Ht} \, .
\end{equation}
For the aristotelian static there is a second central extension given
by
\begin{equation}
  \label{eq:stcent}
[\Pt_{a},\Pt_{b}]=\epsilon_{ab} \Zt_{\Pt} \,.
\end{equation}

The aristotelian static algebra does have a nondegenerate invariant
metric if only the central extension $\Zt_{\Pt}$ is
added.\footnote{This can be understood as a double extension with
  respect to $\Jt$.} It is given by \eqref{eq:invgal} with the
addition of
\begin{align}
  \label{eq:ariststinvm}
  \langle \Pt_{a},\Pt_{b} \rangle &= \delta_{ab}\chi_{\Pt} & \langle \Zt_{\Pt},\Zt_{\Pt}\rangle &= \chi_{\Pt}
\end{align}
under the condition that $\chi_{\Pt}, \chi_{\Ht} \neq 0$.

The torsional static spacetime on the other hand does not have an
invariant metric even if the central extension $\Zt_{\Jt \Ht}$ is
added.

In order to construct a nondegenerate metric for the torsional
aristotelian static algebra we therefore turn again to the double
extension construction. The only permissible choice of subalgebra
$\mathfrak{g}$ compatible with the double extension structure is
$\mathfrak{g}=\{\Pt_a\}$. However, any possible choice of invariant
metric $\langle\Pt_a,\Pt_b\rangle$ is disallowed by the condition
\begin{equation}
  \langle[\Ht,\Pt_a],\Pt_b\rangle+\langle\Pt_a,[\Ht,\Pt_b]\rangle=0 \,.
\end{equation}
Thus again, the only viable double extension is the coadjoint
extension given explicitly by
\begin{subequations}
  \label{eq:torstcoadj}
\begin{align}
    [\Jt,\Pt_a]&=\epsilon_{ab}\Pt_b\,\qquad [\Ht,\Pt_a]=\Pt_a\,\\
  [\Jt,\Pt_a^*]&=\epsilon_{ab}\Pt_b^*\,\qquad [\Ht,\Pt_a^*]=-\Pt_a^*\,\\
  [\Pt_a,\Pt_b^*]&=\epsilon_{ab}\Jt^*+\delta_{ab}\Ht^* \, .
\end{align}
\end{subequations}
The most general invariant metric consists of a part given by
\eqref{eq:invgal} (which is irrelevant for the nondegeneracy), the
usual coadjoint invariant metric
$\langle \Xt_{a}, \Xt_{b}^{*}\rangle= \mu\delta_{ab}$ ($\mu \neq 0$)
for all elements of the original algebra $\Xt_{a}$ and their dual
$\Xt^{*}_{b}$, and a second invariant metric ($\chi \neq 0$) given by
\begin{align}
  \langle \Ht, \Jt^{*} \rangle & = \chi &   \langle \Jt, \Ht^{*} \rangle &= - \chi & \langle \Pt_{a}, \Pt_{b}^{*} \rangle &= \epsilon_{ab} \chi \,.
\end{align}

\subsubsection{$\R \times \Hy^2$ and $\R \times \Sp^2$}
\label{sec:r-times-h2}

Besides \eqref{eq:aristo} the other nonzero commutators for
$\R \times \Hy^2$ (minus sign) and $\R \times \Sp^2$ (plus sign) are
given by
\begin{align}
  [ \Pt_{a} , \Pt_{b}] = \pm \epsilon_{ab} \Jt \,.
\end{align}
We note in passing that the case $\R \times \Sp^2$ corresponds to the
Einstein static universe in 2+1 dimensions.

Since they are a direct sum of a simple and abelian factor it can be
anticipated that they do not posses nontrivial central extensions.
From the above follows that one can construct an invariant metric
given by
\begin{align}\label{eq:RHSdegme}
 \langle \Jt, \Jt \rangle &= \chi & \langle \Pt_{a}, \Pt_{b} \rangle &=\mp \chi \delta_{ab} & \langle \Ht, \Ht \rangle &= \chi_{\Ht}
\end{align}
which is nondegenerate for $\chi, \chi_{\Ht}\neq 0$.

\subsubsection{A24}
\label{sec:a24}

This spacetime is special to $2+1$ dimensions as can be seen by the
commutator
\begin{align}
  [ \Pt_{a}, \Pt_{b}] = \epsilon_{ab} \Ht
\end{align}
which has no higher dimensional analog. It does not permit central
extensions, but has an invariant metric given by
\begin{align}\label{eq:A24metr}
 \langle \Pt, \Pt \rangle &=\mp \chi \delta_{ab} & \langle \Jt, \Ht \rangle &= \chi & \langle \Jt, \Jt \rangle &= \chi_{\Jt} \,.
\end{align}
which is nondegenerate for $\chi\neq 0$.

\section{Generalization to higher dimensions}
\label{sec:gener-high-dimens}

Until this point our discussion was centered around the necessary data
to define Chern--Simons theories based on the kinematical spacetimes
of \cite{Bacry:1968zf,Figueroa-OFarrill:2018ilb}. The double extension
structure turned out to be a useful tool in this endeavor as it allows
us to construct metric Lie algebras as (not necessarily central)
extensions of a given kinematical Lie algebra.

In this section we want to push this program to higher dimensions,
i.e., given a kinematical Lie algebra without nondegenerate metric we
aim to construct a double extension using the smallest number of
additional generators. We will apply this program again to all
kinematical Lie algebras associated to the homogeneous spacetimes of
\cite{Figueroa-OFarrill:2018ilb}.

While extensions of some of these Lie algebras have already been
studied in the literature~\cite{Bonanos:2008kr} we analyze it with
emphasis towards invariant metrics and double extensions. Since the
Lie algebras considered in theories of three-dimensional gravity, at
least in the Chern--Simons formulation, are necessarily metric Lie
algebras our construction can be regarded as a starting point for a
natural generalization of the gauge theory formulation of
three-dimensional gravity to higher dimensions.

Since we work in generic dimensions we employ a somewhat different
notation summarized in Appendix \ref{sec:conventions}. In particular,
the dualization of angular momentum and boosts by the
three-dimensional epsilon tensor is not implied anymore and the Greek
indices $\mu,\nu,\ldots$ run over space and time.

Finally let us emphasize that even though we will refer in the
following to the Lie algebras, we are interested in their
interpretation as spacetimes. Isomorphic Lie algebras
differ due to different choices of associated Klein pairs.

\subsection{(Anti-)de Sitter algebras}
\label{sec:anti-de-sitter}

The (A)dS algebra in $D+1$ dimensions is given in
Table~\ref{tab:adshighermostgen}. We refer to
Appendix~\ref{sec:conventions} for further details on the rescaling
and space-time split of the (A)dS algebra. The generator $\Mt$ is the
single element of the trivial central extension $\mathfrak{u}(1)$ of
the simple (A)dS algebra that can be reabsorbed in the algebra by a
shift of $\Ht$. However, we have included this central extension for
later use in Section \ref{sec:coadj-kinem-algebr}. There we will find
that, as in the three-dimensional case presented in Section
\ref{sec:all-kinem-chern}, the central extension will cease to be
trivial in the nonrelativistic limit. More precisely, after the
nonrelativistic limit the resulting algebra is the Bargmann algebra.

For generic dimension the simple (anti-)de Sitter algebras still
permit an invariant metric which is proportional to the Killing
form. Since the split of the algebra to a semisimple (complex) one
is special to $2+1$ dimensions the invariant metric in generic
dimension is parametrized by the single constant $\chi$ (and an
arbitrary choice of normalization for $\Mt$ given by $\chi_{\Mt}$).
Since the (A)dS algebra already has an invariant metric in any
dimension we do not employ the double extension bootstrap.

\begin{table}[h]
  \centering
$
\begin{array}{l r r}
\toprule %
                                      & [\mathfrak{(a)ds}\oplus \mathfrak{u}(1)]_{\ec,\et,\el} \\ \midrule
  \left[\Jt_{ab},\Jt_{cd}\right]      & 4 \delta_{[a |[c} \Jt_{d]|b]}
                                                                                               \\
  \left[\Jt_{ab},\Bt_{c}\right]       & - 2 \delta_{c[a} \Bt_{b]}
                                                                                               \\
  \left[\Jt_{ab},\Pt_{c}\right]       & - 2 \delta_{c[a} \Pt_{b]}
                                                                                               \\
  \left[\Bt_{a},\Bt_{b}\right]        & \et^{2} \ec^{2}  \Jt_{ab}
                                                                                               \\
  \left[\Bt_{a},\Pt_{b}\right]        & \ec^{2} \delta_{ab} \Ht + \alpha\delta_{ab}\Mt
                                                                                               \\
  \left[\Bt_{a},\Ht\right]            & \et^{2}  \Pt_{a}
                                                                                               \\
  \left[\Pt_{a},\Ht\right]            & \mp \el^{2} \Bt_{a}
                                                                                               \\
  \left[\Pt_{a},\Pt_{b}\right]        & \pm \ec^{2} \el^{2} \Jt_{ab}                           \\ \bottomrule
\end{array}
$
\quad\quad\quad\quad\quad
$
\begin{array}{l r r}
\toprule %
                                      & [\mathfrak{(a)ds}\oplus \mathfrak{u}(1)]_{\ec,\et,\el} \\ \midrule
  \langle \Jt_{ab} , \Jt_{cd} \rangle & 2 \chi \delta_{a[c} \delta_{d]b}
                                                                                               \\
  \langle \Bt_{a} , \Bt_{b} \rangle   & - \chi \et^{2} \ec^{2} \delta_{ab}
                                                                                               \\
  \langle \Ht , \Ht \rangle           & \et^{2} \el^{2} (\pm \chi  +\alpha^{2} \chi_{\Mt})
                                                                                               \\
  \langle \Pt_{a} , \Pt_{b} \rangle   & \mp \chi \ec^{2} \el^{2}  \delta_{ab}
                                                                                               \\
  \langle \Mt , \Mt \rangle           & (\el \et \ec^{2})^{2}\chi_{\Mt}                        \\ \bottomrule
\end{array}
$
\caption{Commutation relations and invariant metric for the trivially
  centrally extended $\mathfrak{ads}$ (upper sign) and $\mathfrak{ds}$
  (lower sign) algebras in dimension $D+1$.}
\label{tab:adshighermostgen}
\end{table}

\subsection{Metric Poincaré algebra and the Maxwell algebra}
\label{sec:poincare-algebra}

As was mentioned at several points already, the Poincaré algebra in
arbitrary dimensions does not permit an invariant bilinear metric. Yet
it is possible to add new generators to the algebra while preserving
most of its commutation relations such that the resulting algebra has
an invariant metric.

The Poincaré algebra is contained in Table \ref{tab:adshighermostgen}
in the $\Lambda\rightarrow 0$ limit setting all other contraction
parameters to unity and quotienting by $\Mt$.

Switching back to relativistic notation for the moment, we find that
the only choice compatible with the double extension structure is
$\mathfrak{g}=\{\Pt_{\mu} \}$, and $\mathfrak{h}=\{\Mt_{\mu\nu}\}$.
The abelian subalgebra generated by $\Pt_\mu$ has a Lorentz-invariant
metric given by the Minkowski metric. With these ingredients we can
construct a double extension by introducing the generators
$\mathfrak{h}^*=\{\Mt_{\mu\nu}^*\}$. The double extension then
dictates the commutation relations in Table \ref{tab:maxwell}.

\begin{table}[h]
  \centering
$
\begin{array}{l r r}
 \toprule %
                                                       & \text{Maxwell}                                     \\ \midrule
 \big[ \Pt_{\mu} , \Pt_{\nu} \big]                     & - \chi_{\Pt} \Mt^{*}_{\mu\nu}                     \\
 \big[ \Mt_{\mu\nu} , \Pt_{\rho}\big]                  & - 2 \eta_{\rho [\mu} \Pt_{\nu]}                      \\
 \big[ \Mt_{\mu\nu} , \Mt_{\rho\lambda}\big]           & \quad 4 \eta_{[\mu|[\rho} \Mt_{\lambda]|\nu]}            \\
 \big[ \Mt_{\mu\nu} , \Mt^{*}_{\rho\lambda}\big]       & 4 \eta_{[\mu|[\rho} \Mt^{*}_{\lambda]|\nu]}        \\  \bottomrule
\end{array}
$
\quad\quad\quad\quad\quad
$
\begin{array}{l r r}
\toprule %
                                                       & \text{Maxwell}                                     \\ \midrule
  \langle \Pt_{\mu}, \Pt_{\nu}\rangle                  & \chi_{\Pt} \eta_{\mu\nu}                           \\
  \langle \Mt_{\mu\nu}, \Mt^{*}_{\rho \lambda} \rangle & \quad 2\eta_{\mu  [\rho} \eta_{\lambda]\nu} \\
  \langle \Mt_{\mu\nu}, \Mt_{\rho \lambda} \rangle     & 2\chi_{\Mt}\,\eta_{\mu  [\rho} \eta_{\lambda]\nu}     \\  \bottomrule
\end{array}
$
\caption{The double extension of the Poincaré algebra is the Maxwell
  algebra. We have lowered the indices of $\Mt^{*\, \mu\nu}$ by
  $\eta_{\mu\nu}$. The invariant metric is nondegenerate
  for $\chi_{\Pt}\neq 0$.}
\label{tab:maxwell}
\end{table}

Remarkably, the double extension leads us to the so called Maxwell
algebra. This long-known algebra describes the symmetries of a
particle in a classical homogeneous electromagnetic
field~\cite{Schrader:1972zd,Bacry:1970ye}. The existence of an
invariant metric for this algebra has been noted in the literature
(see, e.g., \cite{Soroka:2004fj}); here we want to emphasize that the
Maxwell algebra is the most natural extension with an invariant metric
of the Poincaré algebra from the point of view of double extensions.

Our construction also perfectly agrees with Theorem \ref{thm:de},
since we double extend the abelian translations
$\mathfrak{g}=\{\Pt_{\mu} \}$ (which has no factor $\mathfrak{p}$ for
which $H^{1}(\mathfrak{p},\R)$ and $H^{2}(\mathfrak{p},\R)$ are
vanishing) by the simple Lorentz algebra
$\mathfrak{h}=\{\Mt_{\mu\nu}\} \simeq \mathfrak{so}(D-1,1)$.

\subsection{Metric carrollian algebras}
\label{sec:metr-carr-algebr}

Let us now turn to the ultrarelativistic algebras whose commutation
relations are given by the ultrarelativistic limit, $\et \to 0$, of
Table \ref{tab:adshighermostgen} where we ignore the trivial central
extension and set $\ec=1$.

The carrollian algebras do not support a nondegenerate metric in any
dimension other than three, see, e.g.,
\cite{Figueroa-OFarrill:2017tcy}. Due to rotational invariance the
only nonvanishing metric element of $\Ht$ is with itself. But the
invariance condition implies
\begin{equation}
  \label{eq:23}
  0 \eqex \Bt_a\cdot \langle \Ht,\Pt_b\rangle= \langle\Ht,[\Bt_a,\Pt_b]\rangle=\delta_{ab}\langle \Ht,\Ht\rangle\,,
\end{equation}
which renders the metric degenerate. In order to obtain a nondegenerate
metric we will employ again the double extension procedure.

Starting with the flat Carroll-algebra ($\el \to 0$) the first step is to
determine a subalgebra $\mathfrak{g}$ with nondegenerate metric. Due
to the above argument we see that this subalgebra cannot be
$\{\Pt_a,\Ht\}$. The only other option allowed by the structure of the
algebra is $\{\Bt_a,\Ht\}$. However, the flat Carroll algebra is
symmetric under the exchange of $\Bt_a$ and $\Pt_a$ which excludes the
second option, as well. Thus, the Carroll algebra cannot be the
starting point for a double extension apart from the coadjoint
extension (in which $\Ht$ is taken to lie in
$\mathfrak{h}$).\footnote{The Carroll algebra can have a nondegenerate
  metric in three dimensions precisely because it is a coadjoint
  extension.}

Let us turn now to the (A)dS-carrollian algebras. The argument of
equation \eqref{eq:23} still applies, thus $\{\Pt_a,\Ht\}$ is no
option for the subalgebra $\mathfrak{g}$ of the double extension.
Since the (A)dS-Carroll algebra is not symmetric under the exchange of
boosts and spatial translations, one can choose
$\mathfrak{g}=\{\Bt_a,\Ht\}$. A rotational-invariant metric for
$\mathfrak{g}$ is given by
\begin{equation}
  \label{eq:24}
  \langle \Bt_a,\Bt_b\rangle=\chi_{\Bt}\delta_{ab}\quad\qquad \langle\Ht,\Ht\rangle=\chi_{\Ht}\,.
\end{equation}
Invariance under the action of $\Pt_a$ implies the condition
\begin{equation}
  \label{eq:25}
   \chi_{\Ht}=\mp \Lambda^2 \chi_{\Bt}\,.
\end{equation}
Using these as our starting point for a double extension process we
find, introducing the generators
$\mathfrak{h}^*=\{\Jt_{ab}^*,\Pt_a^*\}$, the double extension algebra
$\mathfrak{(a)ds car}^{(E)}_{\el}$ and its invariant metric summarized
in Table \ref{tab:AdSCarrolldoubled}. Here and in the following, we
raise and lower indices of the dual elements using $\delta_{ab}$ and
its inverse. Notice that our extension is precisely of the form
discussed in Theorem \ref{thm:de}, with $\mathfrak{h}$ being simple
and $\mathfrak{g}$ an abelian algebra.

One could have arrived at the double extension of (A)dS-Carroll also
by exchanging the rôle of boosts and spatial translations in the
Maxwell algebra of Table \ref{tab:maxwell}.

The algebra $\mathfrak{(a)ds car}^{(E)}_{\el}$ appears to have gone
unnoticed in the literature. We will add some remarks concerning
possible interpretations of the algebra in the discussion, Section
\ref{sec:discussion}.

\begin{table}[h!]
  \centering
$
\begin{array}{l r r}
\toprule %
                                             & \mathfrak{(a)ds car}^{(E)}_{\el}\\ \midrule
  \left[\Jt_{ab},\Xt_{c}\right] & - 2 \delta_{c[a} \Xt_{b]}
  \\
  \left[\Jt_{ab},\Xt_{cd}\right] & 4 \delta_{[a |[c} \Xt_{d]|b]}
  \\
  \left[\Bt_{a},\Pt_{b}\right] &  \delta_{ab} \Ht
  \\
  \left[\Pt_{a},\Ht\right] & \mp \el^{2} \Bt_{a}
  \\
  \left[\Pt_{a},\Pt_{b}\right] & \pm  \el^{2}  \Jt_{ab}
  \\
  \left[\Pt_{a},\Pt^{*}_{b}\right] & - \Jt^{*}_{ab}
  \\
  \left[\Pt_a,\Jt^*_{bc}\right] & \mp 2 \el^2 \delta_{a[b}\Pt^*_{c]}
  \\
  \left[ \Bt_{a},\Bt_{b}\right] & - \chi_{\Bt}\Jt^*_{ab}
  \\
  \left[ \Bt_{a},\Ht \right] & \pm \Lambda^2\chi_{\Bt}\Pt^{*}_{a} \\
 \bottomrule
\end{array}
$
\quad\quad\quad\quad\quad
$
\begin{array}{l r r}
\toprule %
                                             & \mathfrak{(a)ds car}^{(E)}_{\el} \\ \midrule
  \langle \Bt_{a} , \Bt_{b} \rangle & \chi_{\Bt} \delta_{ab}
  \\
  \langle \Ht , \Ht \rangle &  \mp \Lambda^2\chi_{\Bt}
  \\
  \langle \Jt_{ab} , \Jt^{*}_{cd} \rangle & 2  \delta_{a[c} \delta_{d]b}
  \\
  \langle \Pt_{a} , \Pt^{*}_{b} \rangle &  \delta_{ab}
  \\
  \langle \Jt_{ab} , \Jt_{cd} \rangle & 2 \chi \delta_{a[c} \delta_{d]b}
  \\
  \langle \Pt_{a} , \Pt_{b} \rangle & \mp \chi  \el^{2} \delta_{ab}
  \\
\bottomrule
\end{array}
$
\caption{Double extension based on the AdS-carrollian (upper sign) and
  dS-carrollian (lower sign) algebra. The vector $\Xt_a$ stands for
  $\Bt_a,\Pt_a,\Pt_a^*$; $\Xt_{ab}$ for $\Jt_{ab},\Jt_{ab}^*$. The
  metric is nondegenerate for $\chi_{\Bt}\neq 0$.}
\label{tab:AdSCarrolldoubled}
\end{table}

\subsection{Metric galilean algebras}
\label{sec:metr-galil-algebr}

Let us now turn to the (non-torsional) galilean algebras. These do not
admit an invariant metric for any value of the cosmological constant
$\Lambda$. This is straightforward to check by applying the criterion
\eqref{eq:neccriterion} to the commutation relations obtained in the
limit $c_i\rightarrow 0$ from Table \ref{tab:adshighermostgen},
setting contraction parameter $\tau$ to unity and disregarding the
trivial central extension $\Mt$.

In order to initiate the double extension process we have to choose a
subalgebra $\mathfrak{g}$. The smallest subalgebra that can be
equipped with an $\mathfrak{h}$ invariant metric is the abelian
algebra spanned by boosts and spatial translations. From invariance
with respect to the rotations and nondegeneracy follows that the
invariant metric on $\mathfrak{g}= \{\Bt_{a},\Pt_{a} \}$ is either
\begin{align}
  \label{eq:opt1}
  \langle \Bt_{a}, \Bt_{b} \rangle &= \chi_{\Bt} \delta_{ab}
  &
  \langle \Pt_{a}, \Pt_{b} \rangle &= \chi_{\Pt}  \delta_{ab}
  &
  \chi_{\Bt} \neq 0 \neq \chi_{\Pt}
\end{align}
or
\begin{align}
  \label{eq:opt2}
  \langle \Bt_{a}, \Pt_{b} \rangle &= \chi_{\Bt \Pt}  \delta_{ab} & \chi_{\Bt \Pt}\neq& 0 \,,
\end{align}
or a combination of both.

Checking for invariance under the action of $\Ht$, we find that the
antidiagonal choice \eqref{eq:opt2} is not a viable option. The
diagonal option \eqref{eq:opt1} on the other hand, obeys condition
\eqref{eq:antisymexp} provided that
\begin{align}
  \label{eq:highergalcond}
   \chi_{\Pt} \mp \el^{2}  \chi_{\Bt}=0  \,.
\end{align}

As in the case of the carrollian algebras this means that this
extension exists only for nonvanishing cosmological constant. After
double extending we obtain the metric Lie algebra
$\mathfrak{(a)dsgal}^{(E)}_{\el}$ with commutators and nondegenerate
metric summarized in Table \ref{tab:galcontrhigherD-ext}.
\begin{table}[h]
  \centering
$
\begin{array}{l r r}
\toprule %
                                         & \mathfrak{(a)ds gal}^{(E)}_{\el} \\ \midrule
  \left[\Jt_{ab},\Xt_{cd}\right]         & 4 \delta_{[a |[c} \Xt_{d]|b]}
                                                                            \\
  \left[\Jt_{ab},\Xt_{c}\right]          & - 2 \delta_{c[a} \Xt_{b]}
                                                                            \\
  \left[\Bt_{a},\Ht\right]               & \Pt_{a}
                                                                            \\
  \left[\Pt_{a},\Ht\right]               & \mp \el^{2} \Bt_{a}
                                                                            \\
  \left[\Bt_{a},\Bt_{b}\right]           & - \chi_{\Bt} \Jt^{*}_{ab}
                                                                            \\
  \left[\Pt_{a},\Pt_{b}\right]           & \mp\Lambda^2\chi_{\Bt} \Jt^{*}_{ab}
                                                                            \\
  \left[\Bt_{a},\Pt_{b}\right]           & \mp\Lambda^2\chi_{\Bt}\delta_{ab}\Ht^*
                                                                            \\
  \bottomrule
\end{array}
$
\quad\quad\quad\quad\quad
$
\begin{array}{l r r}
\toprule %
                                         & \mathfrak{(a)ds gal}^{(E)}_{\el} \\ \midrule
   \langle \Bt_{a}, \Bt_{b} \rangle      & \chi_{\Bt} \delta_{ab}
                                                                            \\
  \langle \Pt_{a},\Pt_{b} \rangle        & \pm\Lambda^2\chi_{\Bt}  \delta_{ab}
                                                                            \\
  \langle \Jt_{ab}, \Jt^{*}_{cd} \rangle & 2\delta_{a[c} \delta_{d]b}
                                                                            \\
  \langle \Ht, \Ht^{*} \rangle           & 1
                                                                            \\
  \langle \Ht, \Ht\rangle                & \chi_{\Ht}
                                                                            \\
  \langle \Jt_{ab}, \Jt_{cd} \rangle     & 2\chi_{\Jt} \delta_{a[c} \delta_{d]b}
                                                                            \\
                                                                               
  \bottomrule
\end{array}
$
\caption{Double extension based on the AdS-galilean (upper sign) and
  dS-galilean (lower sign) spacetimes in $D+1$ dimensions and its
  invariant metric. The vector $\Xt_a$ stands for $\Bt_a$,$\Pt_a$; the
  tensor $\Xt_{ab}$ for $\Jt_{ab}$, $\Jt^*_{ab}$. The bilinear form is
  nondegenerate if $\chi_{\Bt}\neq 0$.}
\label{tab:galcontrhigherD-ext}
\end{table}

This can be regarded as a generalization of the three-dimensional
centrally extended (A)dS-Galilei algebra of Table \ref{tab:mostgen}
or, more precisely, a generalization of the double extension procedure
discussed in Section \ref{sec:double-extend-galil}. Since a
non-diagonal metric
$\mu_{\Bt\Pt}$ is not allowed in higher dimensions the linear
transformation in equation \eqref{eq:39} is not necessary.

We want to stress the intriguing observation that the element $\Ht^*$
plays the same rôle as the central extension in the Bargmann algebra,
usually called $\Mt$. We are unaware of previous appearance of this
algebra in the literature. We will postpone further comments and
interpretations to the discussion Section \ref{sec:discussion}.

\subsection{Metric para-galilean and static algebra}
\label{sec:metric-para-galilean}

Even though their Klein pairs and consequently their physics differ,
the para-galilean Lie algebra is isomorphic to the Galilei algebra.
Thus, the discussion of Section \ref{sec:metr-galil-algebr} applies,
we merely need to exchange the boosts and spatial translations. This
means the metric para-galilean algebra is given by the coadjoint
extension, for which we again refer to Section
\ref{sec:coadj-kinem-algebr}.

For completeness we also provide the double extension of the static
algebra in Table \ref{tab:sthigherD-ext}, where
$\mathfrak{g}=\{\Bt_{a}, \Pt_{a}, \Ht\}$ and
$\mathfrak{h}=\{\Jt_{ab}\}$. This is the choice with the smallest
number of additional generators.

\begin{table}[h]
  \centering
$
\begin{array}{l r r}
\toprule %
                                         & \mathfrak{st}^{(E)} \\ \midrule
  \left[\Jt_{ab},\Xt_{cd}\right]         & 4 \delta_{[a |[c} \Xt_{d]|b]}
                                                               \\
  \left[\Jt_{ab},\Xt_{c}\right]          & - 2 \delta_{c[a} \Xt_{b]}
                                                               \\
  \left[\Bt_{a},\Bt_{b}\right]           & - \chi_{\Bt} \Jt^{*}_{ab}
                                                               \\
  \left[\Pt_{a},\Pt_{b}\right]           & - \chi_{\Pt} \Jt^{*}_{ab}
                                                               \\
  \left[\Bt_{a},\Pt_{b}\right]           & - \chi_{\Bt \Pt}\Jt^{*}_{ab}\\
  \bottomrule
\end{array}
$
\quad\quad\quad\quad\quad
$
\begin{array}{l r r}
\toprule %
                                         & \mathfrak{st}^{(E)} \\ \midrule
   \langle \Bt_{a}, \Bt_{b} \rangle      & \chi_{\Bt} \delta_{ab}
                                                               \\
  \langle \Pt_{a},\Pt_{b} \rangle        & \chi_{\Pt}  \delta_{ab}
                                                               \\
  \langle \Bt_{a},\Pt_{b} \rangle        & \chi_{\Bt\Pt}  \delta_{ab}
                                                               \\
  \langle \Ht ,\Ht \rangle               & \chi_{\Ht}
                                                               \\
  \langle \Jt_{ab}, \Jt^{*}_{cd} \rangle & 2 \delta_{a[c} \delta_{d]b}
                                                               \\
  \langle \Jt_{ab}, \Jt_{cd} \rangle     & 2\chi_{\Jt} \delta_{a[c} \delta_{d]b}
                                                               \\
  \bottomrule
\end{array}
$
\caption{Double extension based on the static spacetimes in $D+1$
  dimensions and its invariant metric. The vector $\Xt_a$ stands for
  $\Bt_a$, $\Pt_a$; the tensor $\Xt_{ab}$ for $\Jt_{ab}$, $\Jt^*_{ab}$.
  The bilinear form is nondegenerate if $\chi_{\Ht}$ and
  $(\chi_{\Bt}\chi_{\Pt}-\chi_{\Bt\Pt}^{2})$ are nonzero.}
\label{tab:sthigherD-ext}
\end{table}

\subsection{Non-contracting algebras}
\label{sec:higherdnon-cont}

For completeness, let us also study higher-dimensional generalizations
of the non-contracting algebras of Section \ref{sec:non-contracting}
regarding the existence of invariant metrics. For those algebras
without invariant metric we construct again the double extension with
the smallest number of additional generators.

\subsubsection{Torsional galilean algebras}
\label{sec:torgalmetrichigh}

We saw in Section \ref{sec:tors-galil-theor} that the
three-dimensional torsional galilean theories do not support a
nondegenerate metric, and the only possible double extension is the
coadjoint extension. We will find that this pattern generalizes to
higher dimensions.

The commutation relations of the torsional galilean algebras are given
by
\begin{subequations}
\begin{align}
  \label{eq:26}
  [\Jt_{ab},\Jt_{cd}]& =4 \delta_{[a|[c}\Jt_{b]|d]} & [\Jt_{ab},\Pt_c]&=-2\delta_{c[a}\Pt_{b]} & [\Jt_{ab},\Bt_c]&=-2\delta_{c[a}\Bt_{b]}\,,\\
  [\Ht,\Bt_a]&=-\Pt_a &[\Ht,\Pt_a]&=\alpha \Bt_a+\beta\Pt_a
\end{align}
\end{subequations}
with $\alpha=\gamma,\beta=(1+\gamma), \gamma \in [-1,1)$ for torsional
galilean-dS and $\alpha=(1+\chi^2),\beta=2\chi, \chi>0$ for
torsional galilean-AdS. These algebras do not have an
invariant metric.

The only subalgebras compatible with the double extension structure
are either $\mathfrak{g}=\{\Ht,\Pt_a,\Bt_a\}$ or
$\mathfrak{g}=\{\Pt_a,\Bt_a\}$. However, in both cases the requirement
of invariance under $\Ht$ and $\Jt_a$ leads to a degenerate metric
$\Omega^{\fg}$. As in 2+1 dimensions, the only possibility is
therefore the coadjoint extension, which we refrain from writing, with
its associated nondegenerate metric.

\subsubsection{Carrollian light cone algebra}
\label{sec:carr-light-cone-1}

The carrollian light cone algebra in $D+1$ is given by
\begin{subequations}
\begin{align}
  \label{eq:carrolian_light_cone}
  [\Jt_{ab},\Jt_{cd}] & =4 \delta_{[a|[c}\Jt_{b]|d]} & [\Jt_{ab},\Pt_c] & =-2\delta_{c[a}\Pt_{b]} & [\Jt_{ab},\Bt_c] & =-2\delta_{c[a}\Bt_{b]}\,, \\
  [\Ht,\Bt_a]         & =\Bt_a                       & [\Ht,\Pt_a]      & =-\Pt_a                 & [\Bt_a,\Pt_b]    & =\Ht \,\delta_{ab}+\Jt_{ab}\,.
\end{align}
\end{subequations}
This is isomorphic to $\mathfrak{so}(D+1,1)$. A one-parameter family
of nondegenerate metrics exists due to existence of a nondegenerate Killing form.

\subsubsection{Aristotelian algebras}
\label{sec:arist-algebr}

As in the three-dimensional case, for these algebras the boosts do not
act effectively on the underlying homogeneous space. We have therefore
quotiented by the boosts such that the kinematical algebra is spanned
only by $\{\Jt_{ab},\Pt_a,\Ht\}$, and the kinematical brackets common
to all aristotelian algebras are given by
\begin{equation}
  \label{eq:30}
  [\Jt_{ab},\Jt_{cd}]=4 \delta_{[a|[c}\Jt_{b]|d]}\,\qquad \quad [\Jt_{ab},\Pt_c]=-2\delta_{c[a}\Pt_{b]}\,.
\end{equation}
Let us summarize the results for these algebras:
\begin{itemize}
\item The static aristotelian algebra that is completely defined by
  the commutation relations \eqref{eq:30} has no nondegenerate
  invariant metric. However, one can perform a double extension
  choosing the abelian algebra $\mathfrak{g}=\{\Ht,\Pt_a\}$.
  Introducing the elements $\Jt^*_{ab}$ dual to the rotations
  $\Jt_{ab}$, we obtain a Lie algebra, we denote by
  $\mathfrak{arst}^{(E)}$, with the additional brackets
  \begin{equation}
    \label{eq:31}
    [\Jt_{ab},\Jt^*_{cd}]=4 \delta_{[a|[c}\Jt^*_{b]|d]}\,\qquad \quad [\Pt_a,\Pt_b]=-\mu_{\Pt}\Jt^*_{ab}\,.
  \end{equation}
  and invariant metric
  \begin{align}
        \langle \Ht,\Ht\rangle&=\mu_\Ht &  \langle \Pt_a,\Pt_b\rangle&=\mu_\Pt \delta_{ab} & \langle \Jt_{ab},\Jt^*_{cd}\rangle&=2\delta_{a[b}\delta_{c]d}\,.
  \end{align}
  This is the generalization of the three-dimensional case in Section \ref{sec:static} where
  $\Jt^*$ becomes central.
\item The torsional static aristotelian algebra, having the additional bracket
  \begin{equation}
    \label{eq:33}
    [\Ht,\Pt_a]=\Pt_a\,
  \end{equation}
  has no invariant metric. In contrast to the static case, but again
  similar to all other cases with torsion, the only possible double
  extension is based on the coadjoint algebra since the bracket
  \eqref{eq:33} is incompatible with having a nondegenerate bilinear
  form. We find the additional brackets
  \begin{align}
    \label{eq:34}
     [\Jt_{ab},\Jt^*_{cd}]=4 \delta_{[a|[c}\Jt^*_{b]|d]}\,\qquad [\Pt_a,\Pt^*_b]=-\Jt^*_{ab}\,\qquad [\Ht,\Pt_a^*]=-\Pt_a^*\,\qquad [\Jt_{ab},\Pt_c^*]=-2\delta_{c[a}\Pt^*_{b]}\,
  \end{align}
  and the invariant metric induced from the coadjoint extension structure.
\item The aristotelian algebras $\mathbb{R}\times \mathbb{H}^D$ and $\mathbb{R}\times \mathbb{S}^D$ are defined by
  \begin{equation}
    \label{eq:35}
    [\Pt_a,\Pt_b]=\pm \Jt_{ab}\,,
  \end{equation}
  with the upper sign for the former algebra. These algebras are
  isomorphic to a direct sum of an abelian factor generated by $\Ht$
  and the simple factor $\mathfrak{so}(D,1)$ and $\mathfrak{so}(D+1)$,
  respectively. The kinematical algebra has therefore a two-parameter
  family of invariant metrics, the metric of the simple part being
  proportional to the Killing form.
\end{itemize}

\section{The coadjoint kinematical algebras and their limits}
\label{sec:coadj-kinem-algebr}

In the previous section we examined all higher-dimensional kinematical
spacetimes regarding the existence of an invariant metric for their
kinematical algebra. For those algebras without invariant metric we
found appropriate double extensions that have an invariant metric by
construction. However, the resulting metric Lie algebras do not all
have the same dimension, in particular those of the cube of
Bacry--Lévy-Leblond. If we are interested in considering limits for
these higher-dimensional kinematic algebras with invariant metric and
putative theories based on them, we have to consider the coadjoint
extensions for all algebras. This way we obtain limits for which the
invariant metric stays nondegenerate and which generalizes the outer
cube of Figure \ref{fig:cube} to higher-dimensional kinematical
algebras.

As we mentioned in Section \ref{sec:double-extensions-def} the
coadjoint Lie algebra is $\mathfrak{coad}=D(0,\mathfrak{h})$ defined
on the vector space $\mathfrak{h} \dot +\mathfrak{h}^{*}$ (spanned by
$\Ht_{\alpha}$ and $\Ht^{*\alpha}$, respectively) by
\begin{subequations}
\begin{align}
 [\Ht_{\alpha},\Ht_{\beta}]                   & =f\indices{_{\alpha \beta}^{\gamma}} \Ht_{\gamma}
                             \label{eq:HH1}  \\
 [\Ht_{\alpha},\Ht^{*\beta}]                   & =-f\indices{_{\alpha \gamma}^{\beta}} \Ht^{*\gamma}
                          \label{eq:HdG1}    \\
  [\Ht^{*\alpha},\Ht^{*\beta}]                  & =0
                              \label{eq:HdHd1}
\end{align}
\end{subequations}
and the invariant metric
\begin{align}
  \label{eq:metriccoadjoint}
  \Omega_{ab}^{\mathfrak{coad}}= \bordermatrix{~ & \Ht_{\beta}                       & \Ht^{*\beta} \cr
                             \Ht_{\alpha}     &  h_{\alpha\beta}                   & \delta\indices{_{\alpha}^{\beta}} \cr
                             \Ht^{*\alpha}     &  \delta\indices{^{\alpha}_{\beta}} & 0 \cr} \, .
\end{align}

We take $\mathfrak{h}$ to be the centrally extended (A)dS algebra
of Table \ref{tab:adshighermostgen} with its associated metric $h_{\alpha\beta}$,
that becomes degenerate in the various limits.
Introducing the dual generators of $\fh^*$ and using the coadjoint construction
one arrives at the coadjoint algebra summarized in Table \ref{tab:decfullSTcoad}.
We display only the additional relations
coming from the double extension, all other commutators and matrix
elements are identical to those in Table \ref{tab:adshighermostgen}.

Even though the metric $h_{\alpha\beta}$ becomes degenerate in the IW
contractions the full invariant metric \eqref{eq:metriccoadjoint}
stays nondegenerate. The reason for this is that the rescalings of
equation \eqref{eq:resc} that were used to arrive at Table \ref{tab:adshighermostgen}
are counterbalanced by the inverse rescalings for the dual elements.
This guarantees both nondegeneracy of the metric and regularity of the
contraction~\cite{Prohazka:2017pkc}. This inverse rescaling of the dual elements is already
accounted for when starting directly from the commutation relations in
Table \ref{tab:adshighermostgen}.

\begin{table}[h!]
  \centering
$
\begin{array}{l r r l r }
\toprule %
                                              \multicolumn{5}{c}{D(0,\mathfrak{(a)ds}_{\et,\ec,\el})} \\ \midrule
 \left[\Jt_{ab},\Xt^*_{c}\right]          & - 2 \delta_{c[a} \Xt^*_{b]}        & \quad & \left[\Ht,\Pt_a^*\right]          & \tau^2 \Bt_a^*
                                                                                                      \\
 \left[\Jt_{ab},\Jt^*_{cd}\right]         & 4 \delta_{[a |[c} \Jt^*_{d]|b]}    &       & \left[\Ht,\Bt_a^*\right]          & \mp \Lambda^2 \Pt_a^*
                                                                                                      \\
  \left[\Bt_{a},\Jt^{*}_{bc}\right]       & -2\et^2\ec^2\delta_{a[b}\Bt_{c]}^* &       & \left[\Pt_{a},\Jt^{*}_{bc}\right] &\, \mp 2\Lambda^2\ec^2 \delta_{a[b}\Pt^{*}_{c]}
                                                                                                      \\
  \left[\Bt_{a},\Pt^{*}_{b}\right]        & -\et^{2}\delta_{ab} \Ht^{*}        &       & \left[\Pt_{a},\Bt^{*}_{b}\right]  & \pm \Lambda^2\delta_{ab}\Ht^*
                                                                                                      \\
   \left[\Bt_{a},\Bt^{*}_{b}\right]       & -\Jt^{*}_{ab}                      &       & \left[\Pt_{a},\Pt^{*}_{b}\right]  & -\Jt^{*}_{ab}
                                                                                                      \\
  \left[\Bt_{a},\Ht^{*}\right]            & -\ec^{2} \Pt^{*}_{a}               &       & \left[\Pt_{a},\Ht^{*}\right]      & \ec^{2}  \Bt^{*}_{a}
                                                                                                      \\
  \left[\Bt_{a},\Mt^{*}\right]            & -\alpha \Pt^{*}_{a}                &       & \left[\Pt_{a},\Mt^{*}\right]      & \alpha \Bt_a^*
                                                                                                      \\ \bottomrule
\end{array}
$
\quad\quad\quad\quad\quad
$
\begin{array}{l r r}
\toprule %
                                          & D(0,\mathfrak{(a)ds}_{\et,\ec,\el})                       \\ \midrule
  \langle \Jt_{ab} , \Jt^{*}_{cd} \rangle & 2 \mu^{*} \delta_{a[c} \delta_{d]b}
                                                                                                      \\
  \langle \Bt_{a} , \Bt^{*}_{b} \rangle   & \mu^{*} \delta_{ab}
                                                                                                      \\
  \langle \Ht , \Ht^{*} \rangle           & \mu^{*}   
                                                                                                      \\
  \langle \Pt_{a} , \Pt^{*}_{b} \rangle   & \mu^{*} \delta_{ab}
                                                                                                      \\
  \langle \Mt , \Mt^{*} \rangle           & \mu^{*}
                                                                                                      \\  \bottomrule
\end{array}
$
\caption{Additional commutation relations to Table
  \ref{tab:adshighermostgen} for the coadjoint extension of the AdS (upper sign)
  and dS (lower sign) algebra in $D+1$ dimensions. The vector
  $\Xt^*_a$ stands for $\Bt_a^*$ and $\Pt_a^*$; $\Xt_a$ for $\Bt_a$ and $\Pt_a$. The algebra and its
  metric stay regular for all possible contractions.}
\label{tab:decfullSTcoad}
\end{table}
The invariant metric is nondegenerate for
$\mu^{*} \neq 0$. Due to our construction all limits are
well-defined. Notice that the element $\Mt^*$ dual to $\Mt$ is trivial
as it can be absorbed in a shift of $\Ht^*$. However, in the nonrelativistic
limit it is nontrivial.

These contractions can be seen as the generalization of the cube of
Bacry and Lévy-Leblond~\cite{Bacry:1968zf} to metric Lie algebras and
spacetimes. Indeed, the contractions commute and the invariant metric
stays nondegenerate.

\section{Summary of results}
\label{sec:summary}

This work concerns the classification of gravitational theories in
$2+1$ dimensions and limits of their actions. We analyzed the
underlying structure and the possibility for higher-dimensional
generalizations.

The starting point was the restriction to spatially isotropic
spacetimes, i.e., spacetimes with rotational, spatial and temporal
translations, with boosts (kinematical) and without boosts
(aristotelian)~\cite{Bacry:1968zf,Figueroa-OFarrill:2018ilb}. This is
the natural generalization of the maximally symmetric (A)dS and
Minkowski spacetimes, when the necessity of an invariant tangent space
metric is dropped. The spacetime provides us with a Lie algebra
$\mathfrak{k}$ and a subalgebra $\mathfrak{b}$, the Klein pair
$(\mathfrak{k},\mathfrak{b})$, which fixes the physical interpretation
of the generators of the algebra. This is the starting point for the
gauging of our theories. The gauging provides us with equations of
motion for the vielbein and connection of our gravitational theory. It
does however not provide us necessarily with a well-defined
Chern--Simons action principle, which, among other useful properties,
is the most bullet-proof starting point for the quantization of a
theory.

This led us to analyze the existence of invariant metrics and central
extensions of the underlying Lie algebras, both of which we provide in
full generality (which closes the remaining gaps in the literature).
Since, e.g., the nonrelativistic theories do not provide us with such
a well-defined theory, and motivated by earlier
work~\cite{Papageorgiou:2009zc,Papageorgiou:2010ud,Bergshoeff:2016lwr,Bergshoeff:2016soe,Hartong:2017bwq,Joung:2018frr},
we also fully analyzed the existence of invariant metrics when the
central extensions are added.

For the theories which follow as a limit of (A)dS$_{3}$ gravity we
then show that, when two central extensions are carefully added in the
right way, all limits are well defined also for the action and
organize themselves as a tesseract presented in Figure~\ref{fig:cube}.
The (A)dS-carrollian theories and their limit seem to have been
unnoticed so far.

For the remaining non-contracting cases we introduced the structure
that organizes the non-reductive Lie algebras with invariant metrics,
double extensions. This demystifies the existence of the nondegenerate
``Killing form'' for non-semisimple algebras and made it possible for
us to analyze the detailed structure and paved the way to our main
result: \emph{We provide for every kinematical and aristotelian
  spacetime a well-defined Chern--Simons theory (with the minimal
  number of additional fields).}

We then analyzed if these metric Lie algebras generalize to higher
dimensions (Section~\ref{sec:gener-high-dimens}), recovering the
well-known Maxwell algebra~\cite{Schrader:1972zd,Bacry:1970ye} as a
double extension, and various novel ones, most notably the metric
(A)dS-galilean and (A)dS-carrollian algebras. Finally, we provide a
generalization of the cube to any
dimension~(Section~\ref{sec:coadj-kinem-algebr}).

Before we move on to a more detailed discussion of our results let us
emphasize that our results can be easily adapted to also include the
riemannian ``spacetimes'', see Appendix \ref{sec:riemannian}.

\begin{table}[H]
  \centering
  \caption{Chern--Simons theories that arise from a limit of (A)dS}
  \label{tab:CSlimit}
  \rowcolors{2}{blue!10}{white}
   \resizebox{\textwidth}{!}{
  \begin{tabular}{L L |L C L L | C || L L L}\toprule
                              & \multicolumn{1}{c|}{Limit}     & \multicolumn{4}{c|}{Kinematical CS Theories (Table \ref{tab:kin})} & \multicolumn{1}{c||}{Extended (Table \ref{tab:mostgen})} & \multicolumn{3}{c}{$d > 2+1$}                                                                                                                                                                                     \\
                              & \multicolumn{1}{c|}{($\to 0$)} & \multicolumn{1}{c}{$\langle \,\, , \,\, \rangle$}                  & \multicolumn{1}{c}{\text{ND}}                            &                                            & \multicolumn{1}{c|}{$H^2$} & \multicolumn{1}{c||}{$\Mt$, $\St$} & \text{Sec.}                    &                           & \text{Tab.}                   \\\midrule
      \text{(Anti-)de Sitter} &                                & \mu,\chi                                                           & \yes                                                     & \mugr^{2}  \parmp \el^{2} \muex^{2} \neq 0 &                            & \yes                                       & \ref{sec:anti-de-sitter}       & \text{Simple}             & \ref{tab:adshighermostgen}    \\
      \text{Poincaré}         & \el                            & \mu,\chi                                                           & \yes                                                     & \mu \neq 0                                 &                            & \yes                                       & \ref{sec:poincare-algebra}     & \text{Maxwell}            & \ref{tab:maxwell}             \\
      \text{(A)dS-Carroll}    & \et                            & \mu,\chi                                                           & \yes                                                     & \mu \neq 0                                 &                            & \yes                                       & \ref{sec:metr-carr-algebr}     & \mathfrak{(a)dscar}_\Lambda^{(E)} & \ref{tab:AdSCarrolldoubled}   \\ 
      \text{Carroll}          & \et, \el                       & \mu,\chi                                                           & \yes                                                     & \mu \neq 0                                 & 3, \eqref{eq:Carcentr}     & \yes                                       & \ref{sec:metr-carr-algebr}     & \mathfrak{coad}           & \ref{tab:decfullSTcoad}       \\
      \text{(A)dS-Galilei}    & \ec                            & \mu,\chi,\chi_\Ht                                                  & \no                                                      &                                            & 3, \eqref{eq:galcentr}     & \yes                                       & \ref{sec:metr-galil-algebr}    & \mathfrak{(a)dsgal}_\Lambda^{(E)} & \ref{tab:galcontrhigherD-ext} \\ 
      \text{Galilei}          & \ec, \el                       & \mu,\chi,\chi_\Ht                                                  & \no                                                      &                                            & 3, \eqref{eq:galcentr}     & \yes                                       & \ref{sec:metr-galil-algebr}    & \mathfrak{coad}           & \ref{tab:decfullSTcoad}       \\
      \text{Para-Galilei}     & \et,\ec                        & \mu,\chi,\chi_\Ht                                                  & \no                                                      &                                            & 3, \eqref{eq:pargalcentr}  & \yes                                       & \ref{sec:metric-para-galilean} & \mathfrak{coad}           & \ref{tab:decfullSTcoad}       \\
      \text{Static}           & \et,\ec,\el                    & \mu,\chi,\chi_\Ht                                                  & \no                                                      &                                            & 5, \eqref{eq:statcentr}    & \yes                                       & \ref{sec:metric-para-galilean} & \mathfrak{st}^{(E)}       & \ref{tab:sthigherD-ext}       \\
      \bottomrule
    \end{tabular}
  }
\end{table}
Let us now provide an overview of the theories that admit a limit from
(A)dS$_{3}$ gravity. The vertical rules of Table \ref{tab:CSlimit}
separate the Chern-Simons theories without central extensions (Table
\ref{tab:kin}) from the ones with two, in some cases trivial, central
extension (Table \ref{tab:mostgen}). The last column concerns the
higher-dimensional generalization.

For the nonextended theories the limits of Table \ref{tab:kin} provide
the most general theory except for the freedom to add a
$\langle \Ht , \Ht \rangle = \chi_{\Ht}$ term for the nonrelativistic
theories. In the ND column we have marked the theories that admit a
nondegenerate ($\yes$) or degenerate ($\no$) invariant metric and the
condition for nondegeneracy. This aligns with the inner cube of Figure
\ref{fig:cube}. $H^{2}$ denotes how many nontrivial central extensions
are possible and a link to their commutators. In Section
\ref{sec:central-extensions} the possible invariant metrics for the
centrally extended cases are discussed in detail.

Adding two central extensions ($\Mt$ and $\St$) leads to the
limits of Table \ref{tab:mostgen}, for which all theories admit a
nondegenerate invariant metric as long as \eqref{eq:extnondeg} is
fulfilled. This is also represented in the outer cube of
Figure~\ref{fig:cube}.

In Section \ref{sec:gener-high-dimens} we analyzed if the underlying
metric Lie algebras generalize to higher dimensions. We again describe
how each Lie algebra needs to be altered (if at all) so that we are
able to equip it with an invariant metric. Some correspond precisely
to the higher-dimensional analog of the lower dimensional case ((A)dS,
and (A)dS-Galilei, static) some are a generalization~(Maxwell, (A)dS
Carroll), while for the remaining cases we need to fall back to the
trivial solution of providing them with the coadjoint Lie algebra.

\begin{table}[H]
  \centering
  \caption{Non-contracting Chern--Simons theories}
  \label{tab:CSlnocontr}
  \rowcolors{2}{blue!10}{white}
   \resizebox{\textwidth}{!}{
  \begin{tabular}{L |L L L L L |L ||L L}\toprule
                                                     & \multicolumn{6}{c||}{Non-contracting CS Theories (Section \ref{sec:non-contracting})} & \multicolumn{2}{c}{$d > 2+1$}                                                                                                                                                                                                         \\
                                                     & \text{Sec.}                                                                           & \multicolumn{1}{c}{$\langle \,\, , \,\, \rangle$} & \multicolumn{1}{c}{\text{ND}} &                                    & \multicolumn{1}{c|}{$H^2$} & \multicolumn{1}{c||}{Extended} & \text{Sec.}                 &                  \\\midrule
      \text{Torsional Galilei}                       & \ref{sec:tors-galil-theor}                                                            & \chi_\Jt, \chi_\Ht, \chi_{\Jt \Ht}                & \no                           &                                    & 1                          & \yes, \mathfrak{coad}         & \ref{sec:torgalmetrichigh}  & \mathfrak{coad} \\
      \text{Light cone}                              & \ref{sec:carr-light-cone}                                                             & \mu,\chi                                          & \yes                          & \mu \neq 0 \text{ or } \chi \neq 0 &                            &                                & \ref{sec:carr-light-cone-1} & \text{Simple}    \\ \hline
      \text{Aristotelian static}                     & \ref{sec:static}                                                                      & \chi_\Jt, \chi_\Ht, \chi_{\Jt \Ht}                & \no                           &                                    & 2                          & \yes, 1 \, \text{CE}           & \ref{sec:arist-algebr}      & \mathfrak{arst}^{(E)} \\ 
      \text{Torsional static}                        & \ref{sec:static}                                                                      & \chi_\Jt, \chi_\Ht, \chi_{\Jt \Ht}                & \no                           &                                    & 1                          & \yes, \mathfrak{coad}         & \ref{sec:arist-algebr}      & \mathfrak{coad}      \\
      \text{$\R \times \Hy^2$ and $\R \times \Sp^2$} & \ref{sec:r-times-h2}                                                                  & \chi, \chi_\Ht                                    & \yes                          & \chi, \chi_\Ht \neq 0              &                            &                                & \ref{sec:arist-algebr}      & \text{Reductive}      \\
      \text{A24}                                     & \ref{sec:a24}                                                                         & \chi, \chi_\Jt                                    & \yes                          & \chi \neq 0                        &                            &                                & \multicolumn{2}{c}{\text{Only $2+1$}}               \\
      \bottomrule
    \end{tabular}
  }
\end{table}
Table \ref{tab:CSlnocontr} provides the information concerning the
Chern--Simons theories that do not arise from a limit (middle column)
and their higher-dimensional generalizations (right column).

The Chern--Simons theories for the light cone, $\R \times \Hy^2$,
$\R \times \Sp^2$ and A24 are well-defined from the start ($\yes$),
whereas for the aristotelian static case the addition of one central
extension suffices. For the remaining ones we have to resort to the
coadjoint Lie algebras ($\mathfrak{coadj}$). With $H^{2}$ we again
indicate the existence of nontrivial central extension of which the
precise form can be found in the respective sections.

For the higher dimensional metric Lie algebras the only nontrivial
case is the aristotelian static Lie algebra which is a true
generalization of the lower dimensional one.

More details can be found following the links to the respective
sections (denoted by Sec.).

\section{Discussion}
\label{sec:discussion}

We want to finish by drawing attention to some intriguing aspects of
our classification, mostly focusing on the non-lorentzian theories or
generic features.

\subsection{Three-dimensional kinematical theories}
\label{sec:three-dimensional}

From the $(2+1)$-dimensional point of view it is interesting to
investigate the limit with the AdS/CFT correspondence in mind, i.e.,
the limit from AdS$_{3}$ to AdS-Carroll or AdS-Galilei. One might hope
that the presence of the cosmological constant again translates into
beneficial infrared regulating properties, similarly to AdS. In this
sense these theories also present themselves as interesting testing
ground for the calculations of partition
functions~\cite{Giombi:2008vd,David:2009xg} (see also
\cite{Denef:2009kn,Castro:2017mfj,Porrati:2019knx}, or
\cite{Barnich:2015mui} for the flat case) and for sharpening our
understanding of entanglement entropy, following, e.g.,
\cite{Ammon:2013hba,deBoer:2013vca}, in more general set-ups.

Important for this endeavor would be a deeper understanding of the
solution space, boundary conditions and their asymptotic symmetries,
i.e., hints of possible dual theories. While there has been a lot of
progress for the lorentzian theories, less has been done for their
non-lorentzian cousins, see however
\cite{Bergshoeff:2016soe,Grumiller:2017sjh} for the carrollian and
\cite{Papageorgiou:2009zc,Papageorgiou:2010ud,Bergshoeff:2016lwr,Bergshoeff:2016soe,Hartong:2017bwq}
for the galilean case (see also for related boundary conditions beyond
the kinematical set-up~\cite{Hartong:2017bwq}).

Concerning possible boundary conditions let us add that
following~\cite{Witten:1988hc,Elitzur:1989nr} all models with an
invariant metric can be equipped with boundary conditions leading to
WZW theories on the boundary (for recent work see also
\cite{Grumiller:2016pqb}). Remarkably, similar to the well-known
WZW models based on simple Lie algebras, also the ones based on the
metric Lie algebras permit a Sugawara
construction~\cite{Mohammedi:1993rg} and are therefore conformal field
theories. For the calculation of the central charge in terms of the
level of the WZW model the double extension structure plays a
fundamental rôle~\cite{FigueroaO'Farrill:1994hx} and should be
tractable.

It is interesting that the carrollian Chern--Simons theories are, like
their lorentzian cousins, well-defined from the start. Certainly, the
connection of the ultrarelativistic limit to the strong coupling limit
of general relativity and the Hamiltonian
formulation~\cite{Henneaux:1979vn} deserves further attention.

Since the homogeneous carrollian spacetimes are null surfaces of
Minkowski or (A)dS spaces~\cite{Figueroa-OFarrill:2018ilb} (see also
\cite{Morand:2018tke}), it is tempting to speculate that these
carrollian $2+1$ Chern--Simons theories (in particular the theory
based on the carrollian light cone) are connected to edge states and
BMS symmetries~\cite{Bondi:1962px,Sachs:1962zza} in $3+1$
dimensions. An intriguing hint is that the invariant tensors of the
tangent space of these theories, given by the carrollian degenerate
metric and a carrollian vector field, provide the right carrollian
structure~\cite{Duval:2014uva}. These theories might also yield an
interesting set-up, with manifest carrollian symmetries in the bulk,
for fluid/gravity holographic models~\cite{Campoleoni:2018ltl}.

An obvious generalization is to consider supersymmetric CS theories.
The tools employed here and in \cite{Bergshoeff:2016soe}, should
translate to the supersymmetric case and could clarify the rôle of the
nonrelativistic supergravity theory
of~\cite{Bergshoeff:2016lwr}.\footnote{%
  While the Medina--Revoy theorem has not been proven for Lie
  superalgebras, double extensions still
  work~\cite{FigueroaO'Farrill:1995cy}.}

Furthermore, the double extension structure underlying metric Lie
algebras can be helpful in understanding
Chern--Simons theories based on algebras that do not belong to the
class of kinematical Lie algebras discussed in this work, e.g.,
\cite{Aviles:2018jzw,Concha:2018zeb}.

Similar to the way we classified all three-dimensional Chern--Simons
theories in this work it is possible to carry this framework to two
dimensional dilaton gravity theories. For linear dilaton potentials
these gravity theories can be written as BF theories. Given the
classification of two-dimensional kinematical spacetimes in
\cite{Figueroa-OFarrill:2018ilb} it should be straightforward to
similarly classify all two-dimensional BF theories for these
spacetimes. Since the BF formulation of the Jackiw--Teitelboim model
makes apparent the connection to the Schwarzian action
\cite{Mertens:2018fds,Gonzalez:2018enk}, that rose to prominence due
to its relation to the SYK model \cite{Maldacena:2016hyu}, these
models could have potentially interesting applications regarding
generalizations of the SYK model.

\subsection{Higher dimensions}
\label{sec:higher-dimensions}

Due to their connection to Newton--Cartan/Hořava–Lifshitz
gravity~\cite{Bergshoeff:2016lwr,Hartong:2016yrf} there has been
interest in higher-dimensional generalizations of the double extended
galilean Chern--Simons
theory~\cite{Hansen:2018ofj,Bergshoeff:2018vfn}. It was argued, from
rather different angles, that the actions
of~\cite{Hansen:2018ofj,Bergshoeff:2018vfn}, provide exactly that.

Here we provide another complementary point of view: While the metric
Lie algebra for the flat galilean Lie algebra does \emph{not}
generalize, it is intriguing that it does for \emph{nonvanishing}
cosmological constant to the novel $\mathfrak{(a)dsgal}^{(E)}_{\el}$
algebras (Table \ref{tab:galcontrhigherD-ext}). Let us also emphasize
that the double extension enforces the existence of the central
Bargmann element exactly at the right place for it to be the ``mass''.
We find these observations to be an intriguing and surprising result.

Another place where one might hope for this symmetries to emerge, due
to the fact that they arise in a similar way to the Maxwell algebra, is for
galilean or carrollian electrodynamics in a homogeneous
field~\cite{LeBellac1973}.

Let us finally note that for the flat galilean metric Lie algebra,
like~\cite{Hansen:2018ofj}, we also need to double the dimension of
the algebra. However, as can be easily checked by the criteria of
Section~\ref{sec:central-extensions}, the algebra
in~\cite{Hansen:2018ofj} is not metric, which does not prevent the
authors from writing an action principle. For a natural generalization
of a Chern--Simons theory, i.e., a theory that is invariant under the
full galilean symmetries, our model might be an interesting candidate,
e.g., in the BF theory formulation.

\section*{Acknowledgments} 

SP is grateful to José Figueroa-O'Farrill for insightful discussions
and collaboration on kinematical structures. We thank Glenn Barnich,
Oscar Fuentealba, Joaquim Gomis, Hernán González, Daniel Grumiller,
Marc Henneaux, Victor Lekeu, Wout Merbis, Niels Obers, Gerben Oling,
Jan Rosseel, and Friedrich Schöller for insightful discussions. We
also acknowledge useful feedback from the anonymous referees.

The research of JM and SP is supported by the ERC Advanced Grant
``High-Spin-Grav'' and by FNRS-Belgium (Convention FRFC PDR T.1025.14
and Convention IISN 4.4503.15). JS is supported by the ERC Advanced
Grant GravBHs-692951.

SP would like to express a special thanks to the Mainz Institute for
Theoretical Physics (MITP) for its hospitality and support during the
MITP Topical Workshop ``Applied Newton--Cartan Geometry'' (APPNC2018),
where a preliminary version of our results was presented. The authors
acknowledge support from the Erwin Schrödinger Institute during their
stay at the ``Higher Spins and Holography'' workshop.

SP dedicates this work to Christa Zauner who is now ``auf der anderen
Seite des Weges''.

\appendix

\section{Conventions}
\label{sec:conventions}

For our kinematical Lie algebras and spacetimes we use the following
conventions. Upper case Latin indices denote spacetime indices, while
lower case ones denote spatial indices:
\begin{align}
  \label{eq:indices}
  A &= (\overbrace{0, \underbrace{1, \ldots, D}_{a,b,\ldots}}^{\mu,\nu,\ldots}, \overbrace{D+1}^{\D})
  &
    \eta_{A B}&= \mathrm{diag} (\overbrace{\eta_{00}, \delta_{ab}}^{\eta_{\mu\nu}}, \eta_{\D \D}) \, .
\end{align}
and $\epsilon_{012\cdots D+1}=1$.

Depending on the sign of $\eta_{00}$ we are considering either
lorentzian $(\eta_{00}=-1)$ or Euclidean geometries $(\eta_{00}=+1)$.
Throughout the main text we discuss only the former, but see Appendix
\ref{sec:riemannian}. The sign of $\eta_{\natural\natural}$ determines
the sign of the cosmological constant with anti-de Sitter for
$\eta_{\natural\natural}=-1$ and de Sitter for
$\eta_{\natural\natural}=+1$. We discuss these two cases in
parallel, which leads to $\pm$ and $\mp$ signs, of which the upper
sign always refers to anti-de Sitter and the lower to de Sitter. The
Lie algebras are, if not said otherwise, real and their vector space
elements are denoted in typewriter font, like, e.g., $\Xt$, $\Yt$.

The algebra that leaves $\eta_{AB}$ invariant is then given by
\begin{subequations}
\begin{align}
  [\Mt_{A B},\Mt_{CD}] &= 4 \eta_{[A |[C} \Mt_{D]|B]}=\eta_{AC} \Mt_{DB} - \eta_{AD} \Mt_{CB} - \eta_{BC} \Mt_{DA}  + \eta_{BD} \Mt_{CA}
  \\
  \langle \Mt_{AB}, \Mt_{CD} \rangle &= 2 \mu \eta_{A[C} \eta_{D]B} = \mu (\eta_{AC} \eta_{DB}  - \eta_{AD} \eta_{CB}) 
\end{align}
\end{subequations}
and can also be written, using $\hJ_{\mu\nu}\equiv \Mt_{\mu\nu}$ and
$\hP_{\mu}\equiv \Mt_{\mu\D}$, as
\begin{subequations}
  \label{eq:decMPmet}
\begin{align}
  [\hJ_{\mu\nu},\hJ_{\rho\lambda}] &= 4 \eta_{[\mu |[\rho} \hJ_{\lambda]|\nu]}
  \\
  [\hJ_{\mu\nu},\hP_{\rho }] &= - 2 \eta_{\rho[\mu} \hP_{\nu] }
  \\
  [\hP_{\mu},\hP_{\nu}] &= \pm \hJ_{\mu\nu}
\end{align}
\end{subequations}
with invariant metric
\begin{subequations}
  \label{eq:decMPmet_metric}
\begin{align}
  \langle \hJ_{\mu\nu},\hJ_{\rho\lambda} \rangle &= 2 \mu \eta_{\mu[\rho} \eta_{\lambda]\nu}
  \\
  \langle \hJ_{\mu\nu},\hP_{\rho} \rangle &= 0
  \\
  \langle \hP_{\mu},\hP_{\nu} \rangle &= \mp \mu \eta_{\mu\nu}\,.
\end{align}
\end{subequations}

In order to take the contractions leading to the kinematical algebras
of \cite{Bacry:1968zf}, we find it useful to decompose the generators
further, and introduce the contraction parameters $c_i,\tau$, and
$\Lambda$. We define $\Jt_{ab} \equiv \hJ_{ab}$,
$\Bt_{a}\equiv \hJ_{0a}$, $\Pt_{a} \equiv \hP_{a}$,
$\Ht \equiv \hP_{0}$ and introduce an additional (central) element $\Mt$.
The generators appropriate for the contracted algebras are obtained after
the following isomorphism:
\begin{subequations}
  \label{eq:resc}
\begin{align}
    g_{{\tiny \el,\et,\ec}}\Bt_{a} &= (\et\ec)^{-1}\Bt_{a}
  &
    g_{{\tiny \el,\et,\ec}}\Pt_{a} &= (\el \ec)^{-1}\Pt_{a}\\
    g_{{\tiny \el,\et,\ec}}\Ht &= (\el \et)^{-1}(\Ht+\alpha \ec^{-2}\Mt)
  &
    g_{{\tiny \el,\et,\ec}}\Mt &= (\el \et \ec^{2})^{-1}\Mt\, ,
\end{align}
\end{subequations}
which leads to commutation relations and invariant metric summarized
in Table \ref{tab:adshighermostgen}.


\subsection{Conventions for $2+1$ dimensions}
\label{sec:conv-2+1-dimens}

For $2+1$ dimensions ($D=2$) and restricting to $\eta_{00}=-1$ we have
\begin{align}
  \eta_{\mu\nu}=\mathrm{diag}(-,+,+) \qquad  \eta_{ab}=\delta_{ab}=\mathrm{diag}(+,+)\,.
\end{align}
For the Levi-Civita symbol we define
\begin{align}
 \epsilon_{ab}\equiv \epsilon_{0ab} \quad \qquad \epsilon_{012}=\epsilon_{12}=1  \quad \qquad \epsilon^{ab} = \epsilon_{ab} \,.
\end{align}
We dualize using
\begin{align}
  \hJ_{\mu}= -\frac{1}{2} \epsilon\indices{_{\mu}^{\nu\rho}} \hJ_{\nu\rho}
  \quad
  \Longleftrightarrow
  \quad
  \hJ_{\mu\nu} =  \epsilon\indices{_{\mu\nu}^{\rho}} \hJ_{\rho}
\end{align}
to arrive at~\eqref{eq:AdSdSalgebra}
($\epsilon\indices{_{\mu\nu}^{\rho}}=
\epsilon\indices{_{\mu\nu\lambda}}\eta^{\lambda\rho}$). This implies
that $\Jt_{ab}=-\epsilon_{ab}\Jt^{\mathrm{d}}$ and
$\mathtt{B}_a=\epsilon_{ab}\Bt^{\mathrm{d}}_b$, where on the left-hand
side one has the non-dualized generators. In Sections
\ref{sec:three-dimens-grav},~\ref{sec:double-extensions-1}, and
\ref{sec:non-contracting} we use the dualized operators only and
consistently drop the superscript $\textrm{d}$.

\subsection{Most general centrally extended (A)dS}
\label{sec:most-gener-cent}

In order to arrive at Table \ref{tab:mostgen} we write the commutation
relations \eqref{eq:AdSdSalgebra} using the space-time split
\eqref{eq:split21d} (this is equivalent to Table \ref{tab:kin} setting
all contraction parameters to unity) and add two $\mathfrak{u}(1)$
generators $\Mt$ and $\St$ commuting with all other elements of the
algebra. The invariant bilinear form for the (A)dS part is taken to be
given by \eqref{eq:AdSdSinvm}, and we introduce the additional
parameters $\mu_{\St},\mu_{\Mt},\mu_{\Mt\St}$ for the metric elements
between $\langle \St,\St\rangle$, $\langle \Mt,\Mt\rangle$, and
$\langle \St,\Mt\rangle$, respectively.

The appropriate generators for the contracted algebras are obtained
after applying the transformations
\begin{subequations}
  \begin{align}
    g_{{\tiny \el,\et,\ec}}\Jt           & =\Jt-\beta \ec^2\St                      & g_{{\tiny \el,\et,\ec}}\Bt_{a} & =(\et\ec)^{-1}\Bt_{a} & g_{{\tiny \el,\et,\ec}}\Pt_{a} & = (\el \ec)^{-1}\Pt_{a}
    \\
    g_{{\tiny \el,\et,\ec}}\Ht            & =(\el\et\ec^2)^{-1}(\ec^2\Ht-\alpha \Mt) & g_{{\tiny \el,\et,\ec}}\St     & =\ec^{-2}\St          & g_{{\tiny \el,\et,\ec}}\Mt     & =(\el\et\ec^2)^{-1}\Mt \, ,
  \end{align}
\end{subequations}
while the new metric elements are given by
\begin{align}
  g_{{\tiny \el,\et,\ec}}\mu&= \mugr+\alpha\beta\mu_{\Mt \St} &  g_{{\tiny \el,\et,\ec}}\chi&=\chi+\beta^2\mu_\St 
\end{align}
and
\begin{align}
  g_{{\tiny \el,\et,\ec}}\mu_{\Mt}     & =(\ec \et \el)^{-2} \mu_{\Mt}
                                       & 
  g_{{\tiny \el,\et,\ec}}\mu_{\St}     & =\ec^{-2} \mu_{\St}
                                       & 
  g_{{\tiny \el,\et,\ec}}\mu_{\Mt \St} & =(\et \el\ec^2)^{-1}\mu_{\Mt \St} \, ,
\end{align}
which leads us to  Table \ref{tab:mostgen}.

\subsection{Riemannian spacetimes}
\label{sec:riemannian}

While our analysis was focused on the lorentzian case, it also applies
to riemannian spacetimes. One may reintroduce the sign of the metric
$\eta_{00}$ (see \eqref{eq:indices}) by substituting
$\et^{2} \to - \eta_{00}\et^{2}$ in Table \ref{tab:adshighermostgen}.
With this change one obtains for $\eta_{00} = +1$ the sphere
($\eta_{\D\D} = +1$) and hyperbolic space ($\eta_{\D\D} = -1$), which
both lead to euclidean space in the $\el \to 0$ limit. The analysis in
this work generalizes for the three spaces and mirrors the results of
their lorentzian counterparts. Otherwise, and connected to the fact
that the remaining spacetimes are not lorentzian, the sign $\eta_{00}$
is irrelevant and does not lead to new spacetimes that have not yet
been discussed. For the riemannian spaces the ``boosts'' are compact,
which makes their interpretation as a proper spacetime
questionable~\cite{Bacry:1968zf}.

\section{Kinematical algebras, homogeneous spaces, and Cartan geometries}
\label{sec:kinbackground}

In this appendix we collect relevant background on kinematical
algebras and their relation to homogeneous spacetimes. Furthermore,
introducing the framework of Cartan geometries clarifies the geometric
idea behind the formulation of gravity as a gauge theory. Consult
\cite{Bacry:1968zf,Figueroa-OFarrill:2018ilb,Figueroa-OFarrill:2019sex}
for information on kinematical algebras and homogeneous spacetimes and
\cite{sharpe2000differential,Wise:2006sm} for background (and
hamsters) on Cartan geometries.

\paragraph{Homogeneous spaces.}

The idea of characterizing geometries by their symmetry groups goes
back to Felix Klein and his \emph{Erlangen program}. Consider a
manifold $\mathcal{M}$ and a Lie group $\cK$ acting transitively on
it.\footnote{Let
  $\cK\times \mathcal{M}\rightarrow \mathcal{M}:(g,x)\mapsto gx$ be
  the action of a group on a manifold. The action is called
  \emph{transitive} if for any points $x,y\in \mathcal{M}$ there
  exists an element $g\in \cK$ such that $gx=y$.} Assume further that
there exist ``geometrical structures'' on $\mathcal{M}$ that are left
invariant by a subgroup $\widetilde{\homoH}$ of $\cK$. Then one can
regard the quotient $\cK/\widetilde{\homoH}$ as ``the space of
geometric structures of type $\widetilde{\homoH}$ on $\mathcal{M}$''
(see, e.g., \cite{Wise:2006sm}). In particular, assume the geometrical
structures of interest are the points of the manifold and the subgroup
$\homoH_x$ of $\cK$ leaves a point $x$ invariant. The quotient
$\cK/\homoH_x$ is then $\cK$-equivariantly diffeomorphic to
$\mathcal{M}$. Since a change of base point changes $\homoH_x$ by a
conjugation with an element of $\cK$, the choice of base point is
essentially arbitrary. The manifold $\mathcal{M}$ is therefore
equivalently described in terms of the pair $(\cK,\homoH)$. The space
$\cK/\homoH\simeq \mathcal{M}$ is called \emph{Klein geometry} or
\emph{homogeneous space}.

Disregarding global properties, homogeneous spaces can be
characterized as well by a pair of two Lie algebras $(\fk,\homoh)$,
called \emph{Klein pair} where $\homoh$ is a subalgebra of $\fk$. The
dimension of the homogeneous space is given by the difference in
dimensions of $\fk$ and $\homoh$.

A Klein geometry is called \emph{effective} if $gx=x$ for all
$x\in \mathcal{M}$ implies $g=e$. In other words, the normal subgroup
$\mathcal{N}$ of $\cK$ must be the identity.\footnote{%
  A Klein geometry is called \emph{locally effective} if $\mathcal{N}$
  is a discrete group.} Equivalently, the Lie algebra $\mathfrak{k}$
must not have a non-zero ideal. Any non-effective Klein geometry can
be made effective by replacing $(\cK/\homoH)$ with
$((\cK/\mathcal{N})/(\homoH/\mathcal{N}))$ where $\mathcal{N}$ is the
largest normal subgroup of $\homoH$.

\paragraph{Kinematical spacetimes.}

Not every Klein pair will lead to a Klein geometry that has a sensible
interpretation as a possible stage for physical theories, i.e., as a
\emph{kinematical spacetime}~\cite{Bacry:1968zf}.

In order to have this interpretation it appears reasonable to demand
that the spacetime be smooth, connected and isotropic. The latter
property implies that the group $\cK$ on which the homogeneous space
is based must contain the rotation group as a subgroup.

We employ the following definition \cite{Figueroa-OFarrill:2018ilb}: A
\emph{kinematical Lie algebra} (with $D$-dimensional space isotropy)
is a real Lie algebra $\fk$ with the following two properties:
\begin{itemize}
\item $\fk$ contains a Lie subalgebra
  $\mathfrak{r}\simeq \mathfrak{so}(D)$, the Lie algebra of rotations
  of $D$-dimensional Euclidean space;
\item $\fk$ decomposes as $\fk=\mathfrak{r}\oplus 2V\oplus S$ as a
  representation of $\mathfrak{r}$, where $2V$ are two copies of the
  $D$-dimensional vector irreducible representation of
  $\mathfrak{so}(D)$ and $S$ is the one-dimensional scalar
  representation of $\mathfrak{so}(D)$.
\end{itemize}
A \emph{kinematical Lie group} is a Lie group whose Lie algebra
is a kinematical Lie algebra.

The decomposition of $\fk$ corresponds to the fact that boosts and
spatial translations transform as vectors while time translations
transform as scalars. Notice, however, that the identification of the
generators in $\fk$ with boosts or spatial momenta cannot be made from
the abstract Lie algebra alone. They acquire this interpretation only
in the context of a \emph{homogeneous kinematical spacetime} on which
$\cK$ acts. We quote the definition of
\cite{Figueroa-OFarrill:2018ilb}:

A \emph{homogeneous kinematical spacetime} is a homogeneous space
$\mathcal{M}$ of a kinematical Lie group $\cK$, satisfying the
following properties:
\begin{itemize}
\item $\cM$ is a connected smooth manifold;
\item $\cK$ acts transitively and locally effectively on $\cM$ with
  stabilizer $\homoH$;
\item $\homoH$ is a closed subgroup of $\cK$ whose Lie algebra
  $\homoh$ contains a rotational subalgebra $\mathfrak{so}(D)$
  and decomposes as $\homoh=\mathfrak{r}\oplus V$ as an adjoint
  $\mathfrak{r}-module$, where $V$ is an irreducible $D$-dimensional
  vector representation of $\mathfrak{so}(D)$.
\end{itemize}

The kinematical spacetime is then given by $\cK/\homoH$ and has the
dimension D+1. Its properties are equivalently defined by the
\emph{kinematic Lie pair} $(\fk,\homoh)$.

A classic example of a homogeneous spacetime is $(D+1)$-dimensional
Minkowski space. The Poincaré group $\textrm{ISO}(D,1)$ acts
transitively on this space. Points are left invariant by the Lorentz
group $\textrm{SO}(D,1)$. Thus, Minkowski space can be described as
the homogeneous space $\textrm{ISO}(D,1)/\textrm{SO}(D,1)$. Working on
the infinitesimal level, Minkowski space is characterized by the pair
$(\mathfrak{iso}(D,1),\mathfrak{so}(D,1))$. The Poincaré algebra can
be read off from the $\Lambda\rightarrow 0$ limit of Table
\ref{tab:adshighermostgen}. The Lorentz algebra $\mathfrak{so}(D,1)$
is generated by the set $\{\Bt_a,\Jt_{ab}\}$ which are consequently
interpreted as boosts and rotations, respectively.

The subalgebra $\homoh$ in the kinematic Lie pair thus determines
which generators of $\fk$ are to be interpreted as boosts and
rotations when acting on the homogeneous spacetime. The remaining
generators in $\fk/\homoh$ are then identified with spatial
translations and time translations. In this work we always assume the
following basis for the kinematical algebra~$\fk$:
\begin{equation}
  \label{eq:27}
  \fk=\{\Jt_{ab},\Bt_a,\Pt_a,\Ht\}\,.
\end{equation}

In order to clarify the importance of the choice of subalgebra
$\homoh$ in the kinematical Lie pair, take again the Poincaré algebra
$\fk=\mathfrak{iso}(D,1)$ but quotient now by the subalgebra generated
by $\{\Jt_{ab},\Pt_a\}$, that is not isomorphic to the Lorentz
algebra. Consequently, the resulting homogeneous spacetime is not
Minkowski but the AdS-carrollian spacetime, where now the $\Pt_a$ are
interpreted as boosts and the $\Bt_a$ act as spatial translations on
the spacetime.

In order to make explicit the choice of subalgebra $\homoh$, the basis
\eqref{eq:27} is assumed to be transformed such that
$\homoh=\{\Jt_{ab},\Bt_{a}\}$, i.e., the generators $\Bt_a$ always
denote boosts.

Depending on the form of the kinematical Lie pair a homogeneous
spacetimes may admit additional geometrical structure invariant under
the subgroup $\homoH$. Accordingly, homogeneous spacetimes can be
classified by these invariant structures. For instance, if the
homogeneous spacetime $\cM$ admits an invariant metric on the tangent
space\footnote{This is not to be confused with an invariant metric on
  the kinematical Lie algebra of the homogeneous spacetimes.} it is
called \emph{lorentzian} or \emph{riemannian} depending on the
signature of the metric. If a spacetime admits an invariant one-form
and ``spatial'' co-metric, it is called \emph{galilean}; if it admits
an invariant vector and ``spatial'' metric it is called
\emph{carrollian}. The qualifier ``spatial'' means that the
(co-)metric is degenerate along the invariant (vector) one-form. For
even more details concerning the invariants see Section 6.1.\ in
\cite{Figueroa-OFarrill:2018ilb}.

Additional geometric properties of a homogeneous spacetime can be
derived from an \emph{invariant connection} and its curvature and
torsion. A sufficient criterion for the existence of an invariant
connection is \emph{reductiveness} of the Lie pair. This means that
there exists a split of $(\fk,\homoh)$ as
\begin{equation}
  \label{eq:28}
  \fk=\fm\oplus \homoh
\end{equation}
such that both parts transform as adjoint $\homoH$-modules. This means
that the action of rotations and boosts does not turn translations
into boosts, and vice versa. Notice that the carrollian light cone
algebra discussed in Section \ref{sec:carr-light-cone} is not
reductive and does not have an invariant connection. For further
details we refer to
\cite{Figueroa-OFarrill:2018ilb,Figueroa-OFarrill:2019sex}.

\paragraph{Cartan geometries.}

In the last section we introduced the notion of Klein geometries and
homogeneous spacetimes based on the kinematical Lie pair
$(\fk,\homoh)$. The $\fk$-valued field $A$ appearing in the
Chern--Simons action, or any other ``gauge theory of gravity'' for
that matter, can be qualitatively understood as ``gauging'' of this
underlying Lie pair. This notion is made precise by the concept of
\emph{Cartan geometry}. Roughly speaking, a \emph{Cartan geometry
  modeled on the homogeneous space $\cK/\homoH$} is a manifold $\cM$
on which the linear tangent space is replaced by the homogeneous space
$\cK/\homoH$ at every point. These geometries are therefore
generalizations of riemannian geometries, as summarized in the
following diagram (from \cite{Wise:2006sm,sharpe2000differential}).

\begin{figure}[h]
  \centering
\begin{tikzpicture}

  \node (Klein) [below,align=center] at (0,0) {Klein\\Geometry};
  \node(Cartan) [below,align=center] at (5,0) {Cartan\\Geometry};
  \node  (Riem) [below,align=center] at (5,3) {Riemannian\\Geometry};
  \node(Euc) [below,align=center] at (0,3) {Euclidean\\Geometry};
  \draw[->](Euc) -- node[below,align=center]{\small allow\\\small curvature}(Riem);
  \draw[->] (Klein) -- node[above,align=center]{\small allow\\\small curvature} (Cartan);
  \draw[->] (Euc)--node[left,align=center]{\small generalize\\\small symmetry group}
  (Klein);
  \draw[->] (Riem)-- node[right,align=center]{\small generalize\\ \small tangent space} (Cartan);
\end{tikzpicture}
\caption{Cartan geometry is a generalization of both Riemannian geometry, by allowing the tangent spaces to be different from $\mathbb{R}^n$, and of Klein geometry, by nontrivially gluing together Klein geometries.}
\end{figure}
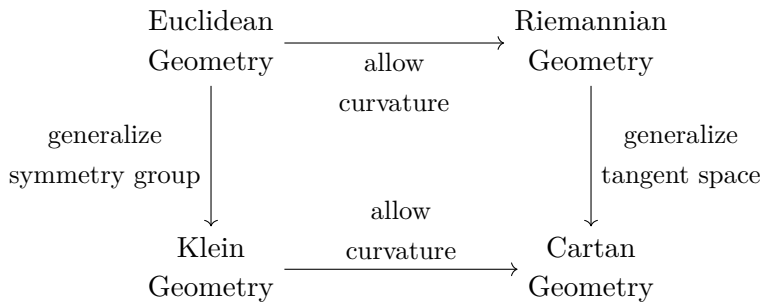

We will refrain from giving the precise definition of a Cartan
geometry modeled on $\cK/\homoH$ which is most naturally phrased as a
principal $\homoH$-bundle $P\rightarrow \cM$ with a $\fk$-valued
1-form on $P$, the \emph{Cartan connection}. It is precisely this
connection that becomes the dynamical quantity in a gauge theory of
gravity. In particular, the Chern--Simons connection \eqref{eq:conn}
is recognized as a Cartan connection $A$.\footnote{Note that this is a
  crucial difference to the concept of an \emph{Ehresmann connection}
  on a principal $\homoH$-bundle over $\cM$ which would be given by a
  $\homoh$-valued one-form. Ehresmann connections are the dynamical
  variables in usual gauge theories as, e.g., Yang--Mills. Thus, gauge
  theories of gravity are a priori different from gauge theories of
  Yang--Mills type. Nevertheless, given some requirements the
  principal bundles with Ehresmann connections and Cartan connections
  can be shown to be equivalent \cite{sharpe2000differential}.}

The curvature $F$ of the Cartan connection, which is a $\fk$-valued
two-form, determines the amount by which the Cartan geometry deviates
from the homogeneous space on which it is modeled. In particular, the
curvature vanishes if the Cartan geometry itself is a homogeneous
space (or a quotient thereof). Even if the Cartan geometry is flat,
i.e., the curvature vanishes, the homogeneous space on which the
geometry is modeled can have curvature. The curvature of the
homogeneous space is given by projecting $F$ to $\homoh$, whereas the
projection of $F$ to $\fk/\homoh$ calculates the $\emph{torsion}$ of
the geometry. Both of these statements are easily checked using the
classic example of $\textrm{AdS}_3$ gravity.

\section{The full action and equations of motion}
\label{sec:full-acti-equat}

While the following formulas might seem intricate let us emphasize
that the presentation is rather economic since we are describing the
theories of Figure \ref{fig:cube} all at once. The most general action
\eqref{eq:4} can be fully decomposed to take the following form
\begin{align}
  S & = \int\left[ 2 \hat{R}\wedge h +2 \left(c_i^2 \mu + \alpha \beta \mu_{\Mt \St}\right) \hat{R}_a \wedge p_a + \hat{\Lambda} +  L(\omega) + \hat{T} \wedge h \pm \Lambda^2\left(c_i^2 \chi + \beta^2 \mu_{\St} \right) \hat{T}_a \wedge p_a  \right. \nonumber \\
    & \quad\quad \left.  + c_{i}^2 L_{M,S} +L_{I} \right] \, ,
\end{align}
where we have dropped various boundary terms and use
\begin{subequations}
\begin{align}
  \hat{R} &=-\mu \, dj+\frac{1}{2}\left(c_{i}^{2}\mu+\alpha \beta \mu_{\Mt\St}\right) \tau^{2}\epsilon_{ab}b_{a}\wedge b_{b}  \\
  \hat{R}_a &=db_a +\epsilon_{ab}b_b\wedge j \\
  \hat{\Lambda} &= \pm\Lambda^{2}\left(c_{i}^{2}\mu+\alpha \beta \mu_{\Mt\St} \right)\epsilon_{ab}p_{a}\wedge p_{b}\wedge h  \\
  L(\omega) &=-\chi j\wedge dj+\tau^{2}\left(c_{i}^{2}\chi+\beta^{2}
              \mu_{\St}\right)\left(b_{a}\wedge db_{a}+\epsilon_{ab}b_{a}\wedge b_{b}\wedge j\right)  \\
  \hat{T}  & =\Big[ \mp \Lambda^2 \tau^2 \chi + \frac{1}{c_i^2} \left( \alpha^2 \mu_{\Mt} \mp \Lambda^2 \tau^2 \beta^2 \mu_{\St} \right) \Big] dh \pm \tau^{2}\Lambda^2 \left(c_{i}^{2}\chi+\beta^{2}\mu_{\St} \right)\epsilon_{ab}b_{a}\wedge p_{b} \\
  \hat{T}_a &= dp_{a}+\tau^{2}\epsilon_{ab}b_{b}\wedge  h +\epsilon_{ab} p_{b}\wedge j  \\
  L_{M,S} &=   \mu_{\Mt} m \wedge dm + \mu_{\St} s \wedge ds +2 \mu_{\Mt \St} m \wedge ds \\
  L_{I} &= 2 \alpha \left(\mu_{\Mt} h\wedge dm + \mu_{\Mt \St}h\wedge ds\right) + 2\beta \left(\mu_{\Mt \St}j\wedge dm+ \mu_{\St} j\wedge ds \right)\, .
\end{align}
\end{subequations}
The equations of motion $F(\Xt_{A}) \equiv \langle F, \Xt_{A}\rangle$
that follow from the variational principle are then given by
\begin{subequations}
\begin{align}
  F(\Ht)&= -2 \mu dh +2 \left(c_{i}^{2}\mu+\alpha \beta \mu_{\Mt\St}\right) \epsilon_{ab}b_a \wedge p_b - 2 \chi dj +\tau^2 \left(c_{i}^2 \chi +\beta^2 \mu_{\St} \epsilon_{ab} b_{a} \wedge b_{b} \right)   \nonumber \\
        &\quad \pm \Lambda^2 \left(c_{i}^{2}\chi+\beta^{2}\mu_{\St} \right) \epsilon_{ab} p_a \wedge p_b +2 \beta \left( \mu_{\Mt \St} dm + \mu_{\St} d s \right) =0  \\
  F(\Pt_{a}) &=  \left(c_{i}^{2}\mu+\alpha \beta \mu_{\Mt\St}\right)\hat{T}_a+  \tau^2 \left(c_{i}^{2}\chi+\beta^{2}\mu_{\St} \right) \left( \hat{R}_a \pm \Lambda^2 \epsilon_{ab} p_b \wedge h\right)=0  \\
  F(\Jt) &= \hat{R} \pm  \Lambda^2 \left(c_{i}^{2}\mu+\alpha \beta \mu_{\Mt\St}\right) \epsilon_{ab}p_a \wedge p_b + 2 \hat{T} +2 \alpha\left( \mu_{\Mt} dm +  \mu_{\Mt \St} d s\right)  =0  \\
  F(\Bt_{a}) &=\pm \Lambda^2 \left(c_{i}^{2}\chi+\beta^{2}\mu_{\St} \right)\hat{T}_a+ \left(c_{i}^{2}\mu+\alpha \beta \mu_{\Mt\St}\right) \left( \hat{R}_a \pm \Lambda^2 \epsilon_{ab} p_b \wedge h\right)=0   \\
  F(\St) &= c_i^2 (\mu_{\St} ds  + \mu_{\Mt \St} dm) +\alpha \mu_{\Mt \St} dh + \beta \mu_{\St} dj   =0  \\
  F(\Mt) &= c_i^2 (\mu_{\Mt} dm +  \mu_{\Mt\St} ds) + \alpha \mu_{\Mt} dh + \beta \mu_{\Mt \St} dj  =0 \,.
\end{align}
\end{subequations}
Under an infinitesimal gauge transformation
$\delta A = d\varepsilon + [A,\varepsilon]$ spanned by a parameter
$\varepsilon = \omega\,\Ht + \lambda^{a}\,\Pt_{a} + \xi
\,\Jt+\sigma^{a}\,\Bt_{a} +\omega^{*}\,\St+\xi^{*} \,\Mt$ valued in
the algebra in Table \ref{tab:mostgen}, the components of the gauge
field transform according to
\begin{subequations}
\begin{align}
  \delta h &= d\omega + c_{i}^2 \epsilon_{ab}(- b_a  \lambda_{b} + p_a  \sigma_b) \\
  \delta p_{a} &= d\lambda_{a} - \epsilon_{ab} j \lambda_{b}+\epsilon_{ab} p_{b}   \xi +\tau^2 \epsilon_{ab} (b_{b}  \omega - h  \sigma_{b}) \\ 
  \delta j &= d\xi - c_{i}^2 \epsilon_{ab} (\tau^2  b_a  \sigma_b \pm \Lambda^2  p_a  \lambda_b )\\
  \delta b_{a} &= d\sigma_{a} - \epsilon_{ab} j  \sigma_{b} +\epsilon_{ab} b_{b}  \xi \pm \Lambda^{2} \epsilon_{ab}(p_{b} \omega - h  \lambda_{b}) \\
  \delta m &= d\xi^{*} + \alpha \epsilon_{ab}  (b_a  \lambda_b - p_a  \sigma_b) \\
  \delta s &= d\omega^{*} + \beta \epsilon_{ab} (\tau^2  b_{a}  \sigma_b \pm \Lambda^2  p_a  \lambda_b)  \, .
\end{align}
\end{subequations}

\bibliographystyle{utphys}
\bibliography{bibl}

\end{document}